\documentclass{aa}  
\def \kosma {KOSMA-$\tau$}
\def \hii {H$_{2}$}
\def \hi {H}
\def \cii {C$^+$}
\def \ci{C{\sc i}}
\def \oi {O{\sc i}}
\def \siii {Si {\sc ii}}

\def \zetacr {$\zeta_\mathrm{CR}$}
\def \tot {\mathrm{tot}}

\usepackage{graphicx}
\usepackage{txfonts}
\usepackage{graphicx}
\usepackage{amsmath}	
\usepackage{booktabs}
\usepackage{hyperref}
\hypersetup{
    colorlinks=true,
    linkcolor=blue,
    urlcolor=blue,
    citecolor=blue
    }
\usepackage{algorithm}
\usepackage{algpseudocode}
\usepackage{subcaption}
\usepackage{mhchem}
\usepackage{multirow}
\usepackage{multicol}
\usepackage{longtable}
\usepackage[dvipsnames]{xcolor}
\usepackage{float}
\begin{document}

\title{Cosmic rays in photodissociation regions I:}
\subtitle{Sensitivity of cosmic ray ionization rate calibrators}
\titlerunning{Cosmic rays in photodissociation regions I}
\authorrunning{Topkaras et al. }

   \author{T. Topkaras$^{1,2}$,\thanks{E-mail: topkaras@ph1.uni-koeln.de}
            V. Ossenkopf-Okada$^{1,2}$,  
            M. R{\"o}llig$^3$, 
            T.~G. Bisbas$^4$,
            B. Gaches$^5$, and
            A. Sinha$^{1,2}$}
    \institute{I. Physikalisches Institut, Universität zu Köln, Zülpicher Str. 77, 50937 Köln, Germany 
    \and{Cluster of Excellence “Our Dynamic Universe” (DYNAVERSE)}
    \and Physikalischer Verein, Robert-Mayer-Straße 2, 60325 Frankfurt am Main, Germany
    \and Research Center for Computational Earth and Space Science, Zhejiang Lab, Hangzhou 311100, China
    \and Faculty of Physics, University of Duisburg-Essen, Lotharstraße 1, D47057 Duisburg, Germany\\    }

   \date{Accepted XXX. Received YYY; in original form ZZZ}

 
  \abstract
   {Low-energy cosmic-rays (CRs) are a key ingredient in photodissociation regions (PDRs). Constraining their impact on photodissociation regions (PDRs) is crucial to understand the underlying chemistry and thermodynamics of those regions.}
   {We investigate how different chemical species correlate with cosmic-ray ionization rates (CRIR, \zetacr) in different environmental conditions, and quantify the impact of CR attenuation on the chemistry of commonly used cosmic-ray ionization rate (CRIR) calibrators.}
   {We implemented a power-law attenuation of \zetacr\, in \kosma\, with a UMIST22-based network of 202 species including ice species. We computed a constant mass grid for 100 M$_{\odot}$, spanning $n_\tot$ = 10–10$^{6}$ cm$^{-3}$, $\chi_0$ = 1–10$^{5}$ (in Draine units) and \zetacr\, = 10$^{-18}$--$10^{-13}$ s$^{-1}$, and quantified the effect of attenuation as an equivalent shift $\Delta$ along the \zetacr\, axis. We measured the CR versus far-ultraviolet (FUV) sensitivity of each species.}
   {CR-heating exceeds photoelectric (PE) heating at low $\chi_0$ and high \zetacr\, values. When CR attenuation is considered, this transition is pushed to \zetacr\, values $\sim$1 dex higher. Identifying species sensitive to CR depends strongly on the environmental conditions. H$_3^+$ is cosmic-ray dominated for over $\sim$ 90\% of the ($\chi_0$, \zetacr) plane at $n_\tot$ = 10$^{5}$ cm$^{-3}$ but only  $\sim$ 47\% at $n_\tot$ = 10$^{3}$ cm$^{-3}$. HCO$^+$ is a clear \zetacr\, tracer for $\chi_0$ $\lesssim$ 10$^3$, while N$_2$H$^+$ only for $\chi_0$ $\lesssim$ 10$^2$. CO and C are insensitive to \zetacr. The effect of attenuation adds an equivalent shift of $\sim$ 0.1-1 dex along the \zetacr\, axis, where a species is CR-dominated. The offset due to attenuation is largest for low \zetacr\, values, and mostly for species formed in the shielded interior.}
   {Estimation of \zetacr\, using a single-species approach carries a systematic error term arising from CR attenuation and an uncertainty based on a far-ultraviolet (FUV)-CR degeneracy. In dense, shielded gas where H$_3^+$, HCO$^+$, and N$_2$H$^+$ are CR-dominated, constant-CR models measure the local attenuated rate, which is approximately a factor of $\sim$ 2-5 below the \zetacr\, incident on the clump. The uncertainty comes from an FUV-CR degeneracy in some species, which has a higher impact in the lower-density gas exposed to FUV. In these environments, a 1 dex uncertainty in $\chi_0$ can add $\sim$ 0.4 dex uncertainty in the inferred \zetacr.}
   \keywords{Astrochemistry --
   Interstellar Medium --
                ISM: cosmic rays --
                ISM: structure --
                photon-dominated region (PDR) --
                methods: numerical}

   \maketitle
%
\section{Introduction}

Photodissociation regions (PDRs) are interstellar medium (ISM) regions whose 
structure and chemistry are strongly influenced by far-ultraviolet 
(FUV, 6 eV $<$ h$\nu$ $<$ 13.6 eV) photon bombardment 
\citep{hollenbach1999photodissociation,wolfire2022photodissociation}. A large fraction of the molecular gas, and by extension almost all the atomic gas, in galaxies resides in PDRs \citep{Hollenbach1997}. Thus, understanding the structure and chemistry of these regions is essential for understanding the processes that regulate star formation in galaxies.

At low visual extinctions, $A_\mathrm{v}$, FUV radiation plays the dominant role in PDRs. As the radiation propagates through the cloud, it excites, dissociates, and ionizes surface species, while species deeper in the cloud (e.g., CO) are shielded from this radiation due to interstellar dust \citep{Glassgold1985}. Thus, the FUV radiation cannot penetrate deep inside the cloud core, leaving a large fraction of chemical processes to be driven by energetic charged particles called cosmic-rays (CRs) \citep{dalgarno2006,Indriolo2013,Gaches2026}. CRs become the primary ionization source at high visual extinctions \citep[$A_\mathrm{v}$  $\gtrsim$ 5 mag;][]{mckee1989,grenier2015}, where their main competitors are X-rays from embedded young stellar objects (YSOs), which penetrate much deeper than FUV photons \citep{krolik1983x,silk1983x}, and shocks. CRs also generate a weak internal FUV field through the excitation of H$_2$ \citep{Prasad1983}, which sustains a low level of photochemistry in regions shielded from external radiation. The ionization is dominated by CR protons with energies between $\sim$1 MeV and 1 GeV, with the low-energy end contributing the most, along with the secondary electrons they produce \citep{dalgarno2006}.

Three usual notations of the cosmic-ray ionization rate (CRIR) $\zeta$ exist in the literature: the primary proton ionization rate, $\zeta_\mathrm{p}$, the total rate of ionization per hydrogen nucleus, $\zeta_\mathrm{H}$, and the total rate of ionization per hydrogen molecule, $\zeta_\mathrm{H_2}$. The first notation is directly related to the density of the CR, while the other two also include ionizations produced by energetic secondary electrons. As shown by \citet{dalgarno1999electron}, the exact ratios among these three quantities depend on both the fractional ionization and the molecular fraction. A simple approximate relation among them is $\zeta_\mathrm{p}$ = $\zeta_\mathrm{H}$ /1.5 = $\zeta_\mathrm{H_2}$/2.3 (i.e., $\zeta_\mathrm{H}$ $\approx$ 0.65 $\zeta_\mathrm{H_2}$) \citep{Glassgold1974}.

Theoretical studies by \citet{Takayanagi1973,Umebayashi1981,Padovani2009} have shown that, although CRs penetrate deeply into molecular clouds, \zetacr\, decreases with increasing column density due to attenuation of the CR flux. This prediction is supported by numerous observational studies that span a wide range of environments, including diffuse molecular clouds, high-mass star-forming regions, low-mass prestellar cores, circumstellar disks, and massive hot cores \citep{caselli1998ionization,derTak_van_dishoeck2000,ceccarelli2004detection,maret2007ionization,hezareh2008simultaneous,shaw2008enhanced,Neufeld2010,IndrioloMcCall2012,ortiz2014ionization,fuente2016ionization,Neufeld2017,Sabatini2020,bialy2022constraining,Sabatini2023,luo2023dependence,luo2023abundance,luo2024new,obolentseva2024reevaluation,Indriolo2026}. The chemical consequences of CR attenuation have been investigated in studies of the early stages of molecular cloud collapse, where the resulting reduction in ionization significantly alters the abundances of key species, including CS, NH$_3$, and solid CO$_2$ \citep{Rimmer2012,Donoghue2022,Gaches2019,Gaches2022a,Latrille2025}. Finally, the diffuse $\zeta_\mathrm{H_2}$ rates derived in most literature works are likely overestimates, as recent re-evaluation of the number densities used for diffuse sight-lines inferred by C$_2$ will reduce them by factors of 2 or higher \citep{Neufeld2024}.

Only recently \citet{Bialy2026} measured $\zeta_\mathrm{H_2}$ directly using CR-excited vibrational H$_2$ emission in the starless core Barnard 68 (B68: with $N_\mathrm{H_2}$ = (1-4) $\times$ 10$^{22}$ cm$^{-2}$ \citep{Nielbock2012,Roy2014}) with the James Webb Space Telescope (\textit{JWST}), following their proposed theoretical approach \citep{Bialy2020}. They derived an average attenuated value of $\zeta_\mathrm{H_2}$ = 1.7 $\pm$ 0.1 $\times$ 10$^{-16}$ s$^{-1}$ at a reference column density of $N_\mathrm{H_2}$ = 3 $\times$ 10$^{21}$ cm$^{-2}$, supporting previous measurements by \citet{Neufeld2024,obolentseva2024reevaluation} and \citet{Indriolo2026}, which reported values in the range $\zeta_\mathrm{H_2}$ = 2.5 $\times$ 10$^{-17}$ - 1.2 $\times$ 10$^{-16}$ s$^{-1}$. However, the value reported by \citet{Bialy2026} lies at the higher end of this range. All previous methods used indirect inference through ions or molecules whose chemistry is connected to CRs (e.g., H$_3^+$). Subsequent work by \citet{Neufeld2026}, using the same observation set, concluded that the measured H$_2$ intensities in the B68 cloud cannot be explained without inferring CR attenuation through a shielding $N_\mathrm{H_2}$ column. These works resulted in higher measurements of $\zeta_\mathrm{H_2}$ (including attenuation), compared to previous methods (e.g., \citealt{IndrioloMcCall2012}). They also determined the attenuation coefficient of penetrating CRs observationally equal to $\sim$ 0.47-0.49 at $\mathcal{R}$ $\sim$ 1.2$\times$10$^{20}$ cm$^{-2}$, where $\mathcal{R}$ is the threshold in column density after which CRs attenuate.

Until recently, most models included CRs as a fixed parameter. Some PDR models have incorporated attenuation in their computations, such as \texttt{UCLCHEM} \citep{Donoghue2022,vermarien2026uclchem}, \texttt{3D-PDR} \citep{Gaches2019,Gaches2022b,Gaches2022a}, \texttt{Nautilus} \citep{Ruaud2016,Clement2023,Wakelam2026}, and \texttt{MEUDON} \citep{Petit2006} (Le Petit, \textit{priv. comm.}). The CR scheme in \kosma\, \citep{rollig2022kosma} is self-consistently coupled to the chemistry, thermal balance, and grain-surface processes. We have added the functionality to optionally attenuate them with depth as a function of the traversed column density, following the power-law prescription described in \citet{padovani2020impact}.

We constructed a 3-dimensional ($n_\tot$=10--$10^{6}$ cm$^{-3}$, $\chi_0$= 1--$10^{5}$ in units of the mean interstellar radiation field \citep{Draine1978}, \zetacr\,= 10$^{-18}$--$10^{-13}$ s$^{-1}$) grid of clump models with a fixed mass of M = 100 M$_{\odot}$, to investigate the CR and CR attenuation impact on the chemistry and structure of PDRs, using the \kosma\, PDR code \citep{rollig2022kosma}. The goal of this study is to identify chemical species whose abundances are sensitive to changes in \zetacr\, under different environmental conditions. We further quantify the impact of CR attenuation using a shift-parameter algorithm to identify the species most strongly affected by it. Furthermore, we develop an analytical framework to estimate the combined uncertainty in the inferred \zetacr\, arising from CR attenuation and the dependence of the chemical abundances on CRs relative to FUV radiation. This paper is structured as follows. In Sect.~\ref{sec:methods} we describe updates to \kosma\, chemistry that we included, in addition to the recent changes presented in \citet{rollig2022kosma} and the attenuation method we implemented in \kosma. Sect.~\ref{sec:results} describes the results of this work, and in Sect.~\ref{sec:discussion} we discuss them. We conclude in Sect.~\ref{sec:conclusions}.

\section{Methods}\label{sec:methods}
\kosma\, \citep{rollig2022kosma} is a well-tested \citep{rollig2007photon} PDR code that employs a spherical cloud geometry or a superposition of many
spherical clumps. Thus, it displays a larger surface-to-volume ratio compared to plane-parallel models using the cloud depth (i.e., \textit{z} = $R_\mathrm{tot}$ - $r$) as a spatial coordinate \citep{Gierens1992,rollig2022kosma}. $R_\mathrm{tot}$ denotes the total radius of the cloud, and $r$ denotes the its local radius from 0 to $R_\mathrm{tot}$. The gas density profile, \textit{n(r)}, used in the code is:

\begin{equation} \label{eq:cloud_density}
 n(r) = n_\tot
\begin{cases}
(\frac{r}{R_\tot})^{-\alpha}, & R_\mathrm{core} < \frac{r}{R_\tot} \leq 1\\
R_\mathrm{core}^{- \alpha} = {\rm const},  &  0 \leq \frac{r}{R_\tot} \leq R_\mathrm{core} ,
\end{cases}
\end{equation}
where $n_\tot$ = $n_\mathrm{H}$ + 2$n_\mathrm{H_2}$ + $n_\mathrm{H^+}$ denotes the total hydrogen gas density on the cloud surface in cm$^{-3}$, $\alpha$ = 1.5 is the density law exponent gradient, and $R_\mathrm{core}$ = 0.2 is the core radius in units of cloud radius. We choose $\alpha$ and $R_\mathrm{core}$ to approximate the structure of a Bonnor-Ebert sphere \citep{Bonnor1956} with a density contrast from edge to center of 11.2 and a clump-averaged density of 1.91 $n_\tot$ \citep{Stoerzer1996}. Performing a spherical integration using the density profile of Eq. \ref{eq:cloud_density}, we can calculate the total hydrogen mass of our clump as:
\begin{equation}
M = 4 \pi m_H n_\tot R_\tot^3 \left[ \frac{R_\mathrm{core}^{3-\alpha}}{3} + \frac{1 - R_\mathrm{core}^{3-\alpha}}{3-\alpha}\right],
\end{equation}
where $m_H$ is the hydrogen mass. The relatively large clump masses of $M = 100\,{\rm M}_{\odot}$ in this study approximate the plane-parallel limit, allowing for comparison with other PDR codes, and ensuring large volumes at sufficiently high column densities for cosmic-ray attenuation to become effective.

\kosma\, solves four independent problems: the local chemical balance, the local level populations of each species, the local energy balance, and finally the radiative transfer equation across the entire spherical geometry of the cloud. To solve the chemical balance, it uses the updated UMIST22 Database for Astrochemistry (\citet{mcelroy1013,millar2024} \footnote{\url{https://umistdatabase.uk}}), which contains 8767 reactions and 737 total species. The total chemical network of \kosma, including the recent addition of surface chemistry species and isotopologues \citep{rollig_oss2013} reported in \citet{rollig2022kosma}, brings the total chemical network to $\sim$39000 reactions and $\sim$1000 species.

For this study, we constructed a reduced chemical network for \kosma\, comprising 202 species and $\sim$ 4500 reactions, including the main isotopologues of C- and O-bearing species and the main species involved in cosmic-ray-initiated chemical pathways (see \citet{Bovino2024} for more details). Species including deuterium (e.g., $\mathrm{HD}$, $\mathrm{H_2D^+}$, $\mathrm{HD_2^+}$, $\mathrm{D_3^+}$, and $\mathrm{N_2D^+}$) were not included. The network includes a sufficiently large number of grain-surface chemistry species, incorporating, e.g., CO freeze-out (denoted as JCO). This effect plays a significant role in high-density regions. Table \ref{tab:surf_chemical_species} in Appendix \ref{sec:surface_species} lists the surface species included in our network, and Table \ref{tab:elem_abund} in Appendix \ref{sec:elem_abundance} shows the initial elemental abundances used by our selected chemical network. Throughout this study, all mentioned abundances are expressed relative to the total hydrogen budget ($n_\tot$ = $n_\mathrm{H}$ + 2$n_\mathrm{H_2}$ + $n_\mathrm{H^+}$, see Eq. \ref{eq:tot_abundance}). Finally, the analysis of the data, for the purposes of this work, was made using the PDR Results Interactive Simulation Mapping \& Analysis toolkit (PRISMA\footnote{\url{https://github.com/theodorostopkaras/prisma-tool}}, for more information see Appendix \ref{sec:prisma}), an open-source PDR post-processing toolkit.

\kosma\ includes several heating and cooling processes. In this study, we will focus on the cosmic-ray heating \citep{Glassgold2012}, and compare it with the other important heating mechanisms. 
CRs can heat the gas by producing hot electrons that collisionally transfer energy to the ISM. The mechanisms of CR and also X-ray heating have been extensively investigated in the literature \citep{dalgarno1999electron,Meijerink2005,Glassgold2012}. Following \citet{Glassgold2012}, CR-heating can be treated as:

\begin{equation}
    \Gamma_\mathrm{CR} = \zeta_\mathrm{H} Q n_\tot \,\, \mathrm{[erg \,s^{-1} \, cm^{-3}]},
    \label{eq:CR_heating}
\end{equation}
where $Q$ denotes the average energy deposited in the gas per ionization. In \kosma\, $Q$ is defined as: 

\begin{equation}\label{eq:qfactor}
    Q(x) = 1.60218 \times 10^{-12} \,\, \mathrm{erg} \frac{4.37}{0.238 + 0.4624e^{-0.42x}},    
\end{equation}
with x = $\mathrm{log_{10}} (\frac{n}{\mathrm{cm^{-3}}})$ \citep{Glassgold2012}.

\subsection{Updated reactions and CR attenuation model} \label{sec:atten_method}
\citet{padovani2018} investigated the ratio of H$_2$ dissociation rate to ionization rate ($ \zeta_\mathrm{diss}/\zeta_\mathrm{ion}$), finding that it should fall between 0.63 and 0.7. This dissociation rate coefficient is determined mainly by the secondary electrons produced during the primary CR ionization process. Recently, \citet{shaw2020} and \citet{sternberg2021} noted that the value of the dissociation of molecular hydrogen adopted in the UMIST database does not reflect this ratio. Using the new re-calibrations and computations of this ratio, the reaction rate for direct CR dissociation for the reaction:

\begin{align}
    \rm H_2 + p \rightarrow &\ \rm H + H + p\tag{R1} \label{R1}, 
\end{align}
where p is a CR proton, is revised to be 9$\times$10$^{-18}$ s$^{-1}$ instead of the previously reported value of 1.3$\times$10$^{-18}$ s$^{-1}$ \citep{millar2024}. This change has major consequences for the evolution of the chemical network because atomic hydrogen remains a key reactive species in dense molecular clouds ($N_\mathrm{H_2}$ $\geq$ 10$^{21}$ cm$^{-2}$), where UV photodissociation is suppressed by H$_2$ self-shielding and CR-induced dissociation becomes increasingly important. Finally, \citet{Padovani2026} showed that the contribution of low-energy CR protons to proton-induced H$_2$ excitation can be comparable to that of secondary electrons.

A recent study by \citet{Luo2026} further updated the value deduced by \citet{padovani2018}, to $\zeta_\mathrm{diss}$ = 0.589$\zeta_\mathrm{ion}$ by including the contributions of dissociative ionization and dissociative electron capture. In their study, they also updated the total destruction rate of H$_2$ by CRs as:
\begin{equation}\label{eq:luo2026_new_dest_rate}
    \zeta^\mathrm{d} = \zeta^\mathrm{ion} + \zeta^\mathrm{diss} + \zeta^\mathrm{cr-uv,ion} + \zeta^\mathrm{cr-uv,diss},
\end{equation}
with $\zeta^\mathrm{cr-uv,ion}$ and $\zeta^\mathrm{cr-uv,diss}$ defining the CR-induced photoionization and photodissociation of H$_2$, respectively. Accounting for all processes, they give a value of $\zeta^\mathrm{d}$ = 2.47 $\zeta^\mathrm{ion}$ for the ratio of the total cosmic-ray destruction rate to the ionization rate. \kosma\, includes CR-induced photodissociation and photoionization of H$_2$ self-consistently through the internal UV radiative transfer, resulting in a somewhat lower efficiency of those processes (R{\"o}llig et al. \textit{in prep}). Thus, for \kosma\, we have $\zeta^\mathrm{d}$ = 1.64 $\zeta^\mathrm{ion}$, instead of the 2.47 found by \citet{Luo2026}.

In \kosma\, the the CRIR is denoted as \zetacr, and sets every CR reaction rate as $k_\mathrm{CPR}\,\, [\mathrm{s}^{-1}] = \alpha_{\rm diss}\cdot\zeta_{\rm CR}/\zeta_0$, where $\alpha_\mathrm{diss}$ are the UMIST22 rate coefficients and $\zeta_0$ = $1.3 \times 10^{-17}$ s$^{-1}$ is the reference CR rate adopted in the network. Summing the ionizing channels of H$_2$ (H$_2^+$ + e$^-$, H$^+$ + H + e$^-$, H$^+$ + H$^-$) gives the H$_2$ ionization rate applied in \kosma\, as $\zeta_\mathrm{H_2}$ = 0.945\zetacr. Cosmic-ray attenuation was incorporated in the model using the prescription described by \citet{Padovani2009, padovani2020impact}, approximated by a simple power-law dependence on the column density as:

\begin{equation} \label{eq:pad2009method1}
\zeta_\mathrm{CR}\,\, [s^{-1}] = \zeta_\mathrm{CR,0}
\begin{cases}
 \left(\frac{N_\mathrm{eff}}{\mathcal{R}}\right)^{- \alpha}, & N_\mathrm{eff}  \geq \mathcal{R}\\
1,  &  N_\mathrm{eff} < \mathcal{R}.
\end{cases}
\end{equation}

\begin{figure}[ht]
\centering
\includegraphics[width = 3.5in]{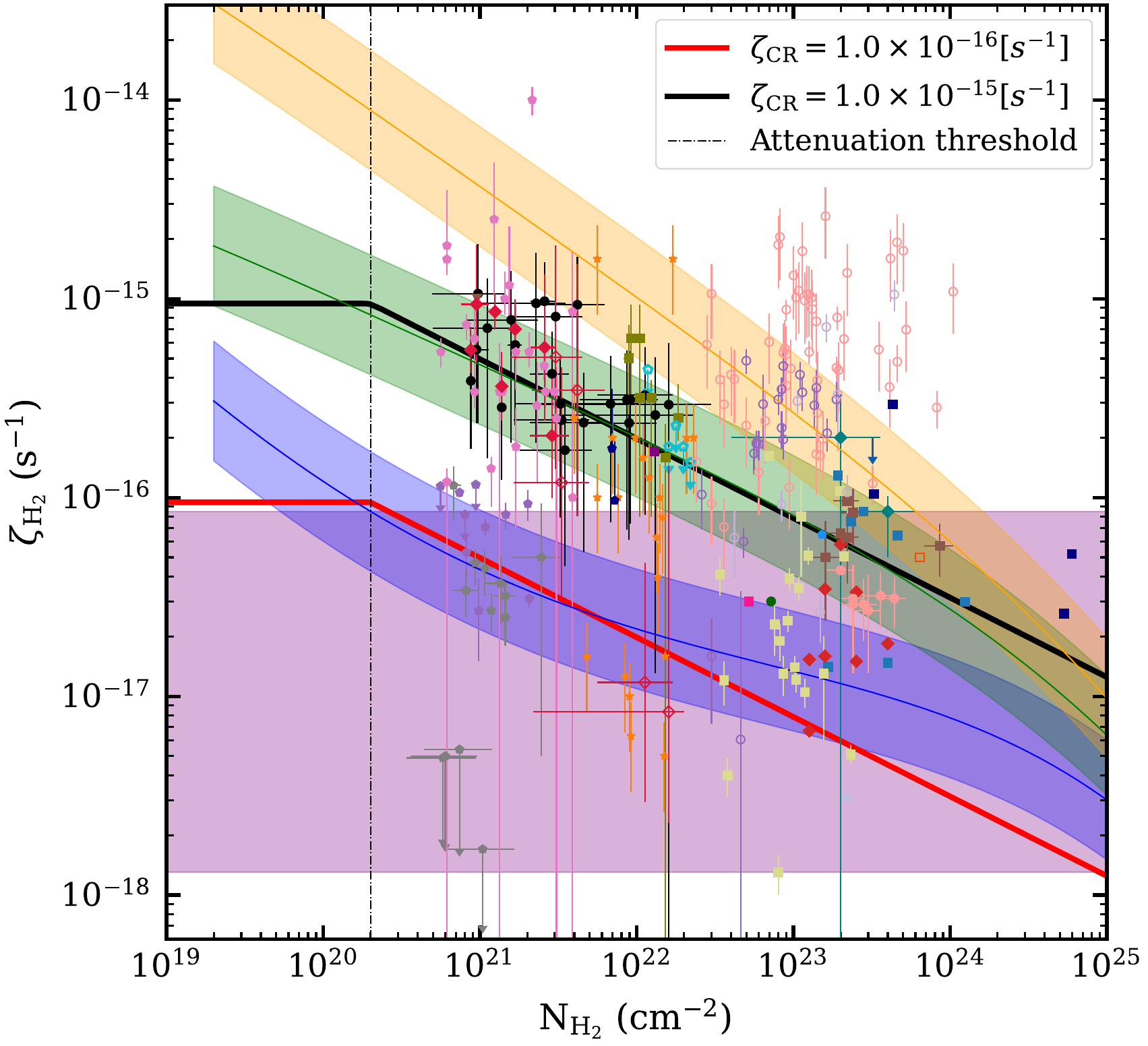}
\caption{Cosmic-ray ionization rate of molecular hydrogen, $\zeta_\mathrm{H_2}$, as a function of the molecular hydrogen column density, $N_\mathrm{H_2}$. Models $\mathcal{L}$, $\mathcal{H}$, and $\mathcal{U}$ presented in \citet{padovani2024ultraviolet} are included with blue, green, and orange continuous lines, respectively, with a shaded area denoting a factor of 2 error from the model. The black and red curves show the attenuated \kosma\ models with \zetacr\, = 10$^{-15}$ s$^{-1}$ and 10$^{-16}$ s$^{-1}$, respectively, using an attenuation coefficient $\alpha$ = 0.4. The vertical dashed-dotted line denotes the attenuation limit at $N_\mathrm{eff}$ = 10$^{20}$ cm$^{-2}$. The observational estimates included on the figure are as follows: black circles \citep{IndrioloMcCall2012}, blue squares \citep{derTak_van_dishoeck2000}, x symbol \citep{hezareh2008simultaneous}, orange stars \citep{caselli1998ionization}, red diamonds \citep{Sabatini2020}, light pink circles and brown squares \citep{Sabatini2023}, purple pentagons \citep{obolentseva2024reevaluation}, pink pentagons \citep{luo2023abundance}, empty cyan pentagons \citep{bialy2022constraining}, down pointing brown triangles \citep{shaw2008enhanced}, green circles \citep{maret2007ionization}, blue pentagons \citep{fuente2016ionization}, teal diamonds \citep{ortiz2014ionization}, empty red squares \citep{ceccarelli2004detection}, navy squares \citep{Barger_2020}, empty pink, purple and light purple circles \citep{luo2024new}, olive squares \citep{luo2023dependence}, gray polygons \citet{Indriolo2026}, navy polygons \citep{Neufeld2010}, crimson diamonds \citet{Neufeld2017}, and purple square \citep{Bialy2026}. Finally, the purple shaded area shows the values of $\zeta_\mathrm{H_2}$ reported by \citet{Redaelli2025}.}
\label{fig:crir_atten}
\end{figure}

    In the above equation, $\zeta_\mathrm{CR,0}$ is the constant CR flux of the model, $N_\mathrm{eff}$ = 0.5($N_\mathrm{H}$ + 2$N_\mathrm{H_2}$) is the effective total column density we need to consider to be consistent with the adopted method from \citet{padovani2020impact}, and $\alpha$ is the attenuation coefficient that defines the steepness of the attenuation. In addition, $\mathcal{R}$ defines the threshold for which the attenuation effectively starts (i.e., for  $N_\mathrm{eff}$ $>$ $\mathcal{R}$, $\zeta_\mathrm{CR,0}$ will be attenuated). Equation \ref{eq:pad2009method1} is valid for $2\times10^{19}\leq {\rm N(H_2)}\leq 2\times10^{25}\,{\rm cm}^{-2}$ . For $N_\mathrm{H_2}$ $\geq$ 2 $\times$ 10$^{25}$ cm$^{-2}$ the \zetacr\, depends on the high-energy-tail CR particles of the incident spectrum. It attenuates exponentially with the column density of the traversed matter. Because this regime is beyond the scope of this work, we will not examine this description. Finally, \citet{Adams2014} showed that massive stars can deliver short-lived radioactive decay nuclei, contributing to both heating and ionization of their surrounding environment. This sets a low limit for the ionization rate, especially in regions where $\zeta_\mathrm{CR}$ is attenuated, of $\sim$ (1-5) $\times$ 10$^{-19}$ s$^{-1}$.

Figure \ref{fig:crir_atten} shows, with solid red and black lines, the implemented power-law attenuation of our model for \zetacr\, = 10$^{-16}$ s$^{-1}$ and 10$^{-15}$ s$^{-1}$, respectively. The power-law attenuation coefficient is taken equal to $\alpha$ = 0.4, following an approximate value based on \citet{Padovani2009} and recent observational data \citep{Neufeld2017,Neufeld2026}. The attenuation threshold, $\mathcal{R}$, is shown with a vertical dashed-dotted line at $N_\mathrm{H_2}$ = 2 $\times$ 10$^{20}$ $\rm cm^{-2}$. The shaded purple region indicates the range of $\zeta_\mathrm{H_2}$ =  (1.3 $\times$ 10$^{-18}$ s$^{-1}$ - 8.5 $\times$ 10$^{-17}$ s$^{-1}$) reported by \citet{Redaelli2025}. The blue, green, and orange curves show the models $\mathcal{L}$, $\mathcal{H}$, and $\mathcal{U}$, respectively, from \citet{padovani2024ultraviolet}, with the shaded regions corresponding to a factor of 2 uncertainty. Model $\mathcal{L}$ corresponds to a lower-energy extrapolation of Voyager 1 observations, while Model $\mathcal{H}$ displays an enhanced flux of low-energy protons with respect to Voyager 1 data and best reproduces the AMS-02 high-energy CR spectrum \citep{aguilar2014electron,aguilar2015precision,Cummings2016}. Finally, Model $\mathcal{U}$ was recently introduced by \citet{padovani2024ultraviolet} to best reproduce the ionization rate values that match the upper envelope of the available observational estimates \citep{Padovani2022}.

\subsection{Regime binning of the parameter space} \label{sec:binning_method}
Several quantities that will be discussed in this work, such as the CR dominance tracers (see Sect. \ref{sec:fuv_vs_cr}) and the attenuation shift $\Delta$ (see Sect. \ref{sec:attenuation_effect}), vary strongly across our ($n_\tot$, $\chi_0$, \zetacr) parameter space. A single grid slice for one parameter can adequately represent specific environmental conditions. However, to derive a statistical conclusion, we need to use the full parameter space. Therefore, we divide each axis into three logarithmically spaced intervals (i.e., low, mid, and high), resulting in 27 bins corresponding to all combinations of grid regimes, as shown in Table \ref{tab:bins}. As all models have the same total clump mass of 100 $M_{\odot}$, the total column density of hydrogen nuclei, $N_\tot$, from the surface to the center of the clump is not an independent quantity. Table \ref{tab:bins} presents the $N_\tot$ range corresponding to each density bin.

\begin{table}[ht]
\centering
\caption{Logarithmic bin splitting of the three-dimensional ($n_\tot$, $\chi_0$, \zetacr) parameter space.}
\resizebox{\columnwidth}{!}{%
\begin{tabular}{@{}lccc@{}}
\toprule
Regime                          & Low             & Mid              & High           \\ \midrule
$\log_{10}$$n_\tot$ [cm$^{-3}$] & 1-2.67          & 2.67-4.33        & 4.33-6 \\
$\log_{10}$$\chi_0$ [Draine]    & 0-1.67          & 1.67-3.33        & 3.33-5 \\
$\log_{10}$\zetacr\, [s$^{-1}$] & (-18) - (-16.33) & (-16.33) - (-14.67)  & (-14.67) - (-13 )\\ 
$\log_{10}$N$_\tot$ [cm$^{-2}$] & 20.73-21.84     & 21.84-22.95      & 22.95-24.06 \\ \bottomrule
\end{tabular}
\label{tab:bins}
}
\end{table}

\section{Results}\label{sec:results}

\subsection{Gas heating and cooling mechanisms}
To compare the heating and cooling processes included in \kosma, we calculate the clump-integrated heating and cooling rates as:

\begin{equation}\label{eq:clump_heat_cool}
\Gamma_\mathrm{clump} \,\, \mathrm{or} \,\, \Lambda_\mathrm{clump} = \int_0^R 4\pi r^2 \Gamma (r) \,\,(\mathrm{or} \,\,\Lambda (r))\, dr
\end{equation}
where $\Gamma(r)$ and $\Lambda(r)$ denote the local volumetric heating and cooling rates in units of [erg cm$^{-3}$ s$^{-1}$], respectively, for each process at each radial position within the clump.

\subsubsection{Heating mechanisms}

\begin{figure*}
\centering
\includegraphics[width = 7in]{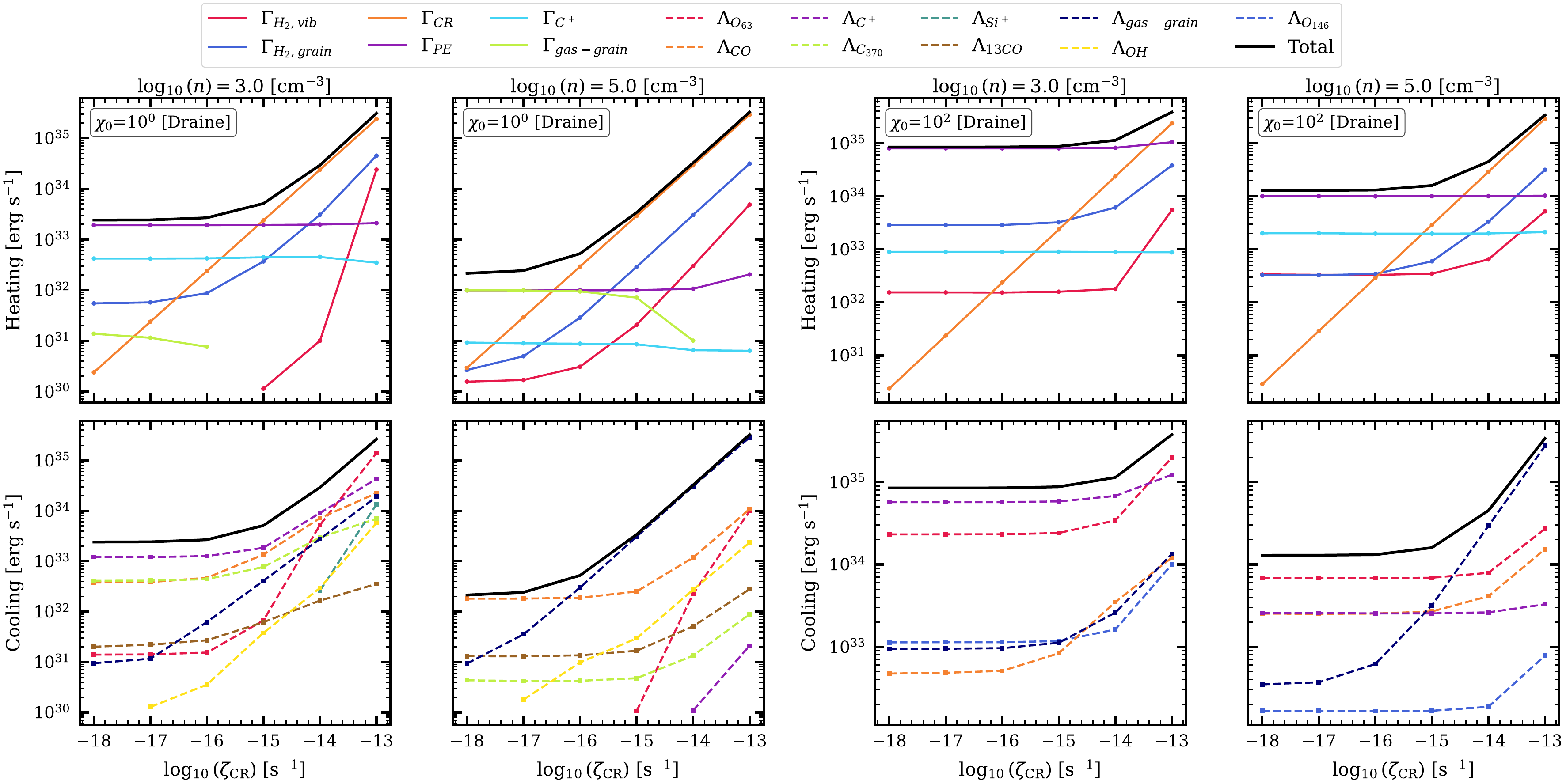}
\caption{Clump-integrated heating and cooling contributions for models with gas densities of n=10$^{3}$ cm$^{-3}$ and n=10$^{5}$ cm$^{-3}$ and FUV radiation fields of $\chi_0$ = 1 and $\chi_0$ = 10$^{2}$ (in Draine units), assuming a constant \zetacr. Colored solid lines denote the individual heating processes, while dashed lines represent the major cooling processes. The total heating and cooling rates are shown by the solid black line in each panel. For the heating process regarding H$_2$ we include the H$_2$ vibrational de-excitation ($\Gamma_\mathrm{H_2,vib}$), and the H$_2$ formation on grains ($\Gamma_\mathrm{H_2,grain}$).}
\label{fig:heating_cooling_mech}
\end{figure*}

The upper panels of Fig. \ref{fig:heating_cooling_mech} show the $\Gamma_\mathrm{clump}$ calculated using Eq. \ref{eq:clump_heat_cool} for two density regimes, $n_\tot$ = 10$^{3}$ and 10$^{5}$ cm$^{-3}$, as a function of \zetacr. The two left panels correspond to $\chi_0$ = 1, while the two right panels correspond to $\chi_0$ = 100. We show only the constant CR models ($\alpha$ = 0) and include only those mechanisms that rank among the three dominant contributors to the heating budget, at some point in the grid. Mechanisms with smaller contributions are omitted from the figure. The corresponding figure for models with attenuation is given in the Appendix in Fig.~\ref{fig:heating_cooling_mech_atten}.

The clump-integrated CR-heating rate profile ($\Gamma_\mathrm{CR}$) is nearly identical in all panels of Fig. \ref{fig:heating_cooling_mech}. This happens because at fixed clump mass it scales with \zetacr\, and $n_\tot$ (Eq.~\ref{eq:CR_heating}), independently of $\chi_0$, and Q varies only from 11.8 to 14.8 eV between $n_\tot$ = 10$^3$ and 10$^5$ cm$^{-3}$ (Eq. \ref{eq:qfactor}). $\Gamma_\mathrm{CR}$ therefore acts as a fixed reference against which the FUV-driven channels compete, and the ionization rate at which CRs take over the energy budget directly measures the strength of the FUV heating. $\Gamma_\mathrm{PE}$ is confined to the FUV-illuminated surface, and dominates at low \zetacr\, in all models. $\Gamma_\mathrm{PE}$ increases with $\chi_0$ and decreases with $n_\tot$ as the clump surface shrinks. Consequently, $\Gamma_\mathrm{CR}$ becomes dominant at \zetacr\, $\gtrsim$ 10$^{-15}$ s$^{-1}$ for $\chi_0=1$ and $n_\tot$ = 10$^{3}$ cm$^{-3}$, and at \zetacr\, $\gtrsim$ 10$^{-17}$ at $\chi_0=1$ and $n_\tot$ = 10$^{5}$ cm$^{-3}$. For the stronger FUV field, $\chi_0=10^{2}$, this transition occurs at higher ionization rates: \zetacr\, $\gtrsim$ 10$^{-14}$ and 10$^{-15}$ s$^{-1}$ for $n_\tot$ = 10$^{3}$ and 10$^{5}$ cm$^{-3}$, respectively.

In steady state, H$_2$ formation heating ($\Gamma_\mathrm{H_2,grain}$) traces the H$_2$ destruction rate, which at high \zetacr\, is dominated by CRs throughout the clump volume rather than by FUV photodissociation at the surface. H$_2$ formation heating therefore increases linearly with \zetacr\, and converges to the same value in all models, contributing $\sim$10\% of the total heating. At low \zetacr\, it competes with \cii\, heating ($\Gamma_\mathrm{C^+}$), which dominates where the low $\chi_0/n_\tot$ ratio keeps the atomic hydrogen column small \citep{Sternberg2014}. Heating by H$_2^*$ de-excitation ($\Gamma_\mathrm{H_2,vib}$) contributes only for \zetacr\, $\gtrsim$ 10$^{-14}$ s$^{-1}$.

Including cosmic-ray attenuation reduces the contribution of $\Gamma_\mathrm{CR}$ and shifts the transition to CR-dominated heating towards higher \zetacr, with the effect depending on density and FUV field. The corresponding changes in individual heating and cooling processes are discussed in the Appendix \ref{sec:remaing_heating_cooling}.

\begin{figure*}
\centering
\begin{subfigure}[t]{0.48\textwidth}
    \centering
    \includegraphics[width=0.48\linewidth]{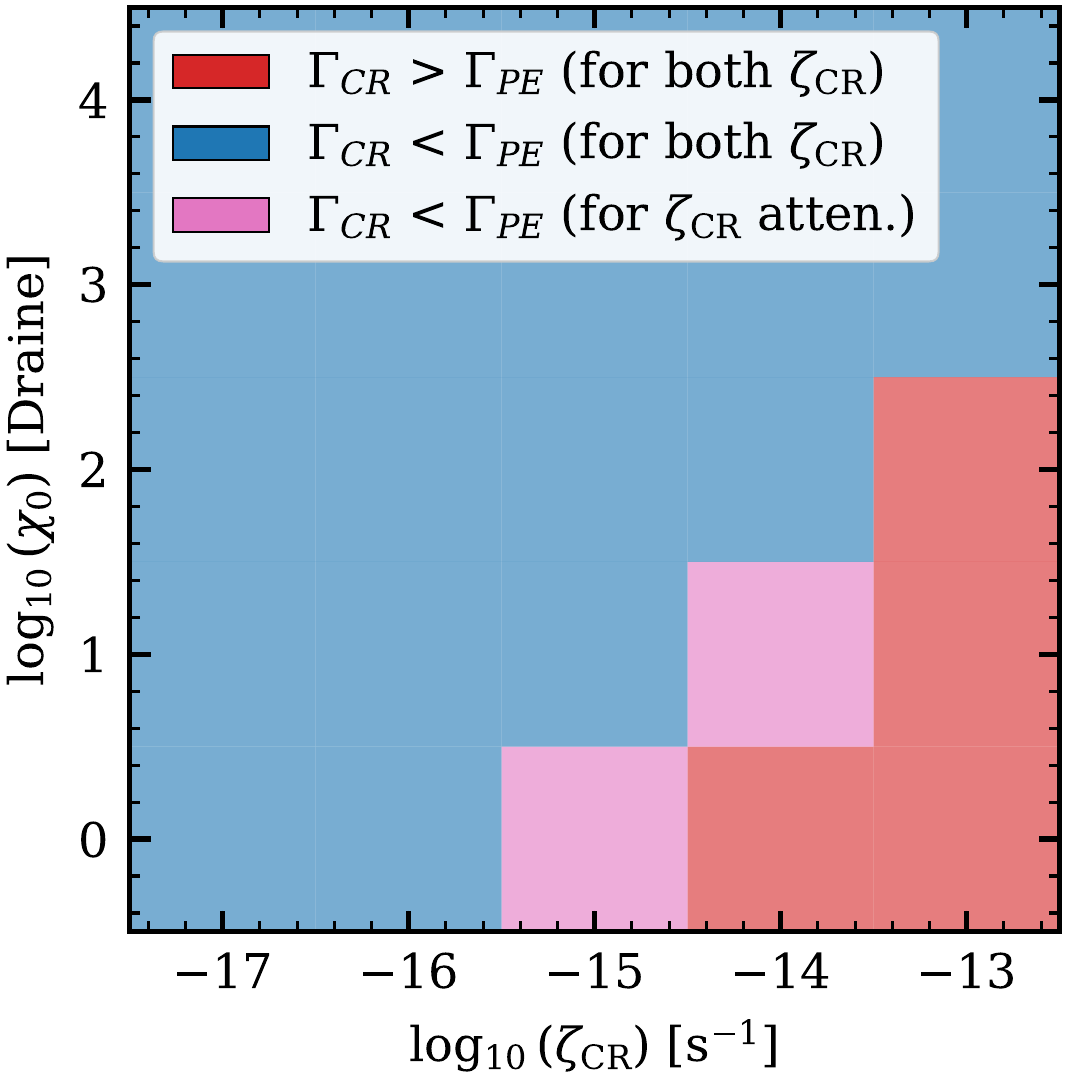}
    \hfill
    \includegraphics[width=0.48\linewidth]{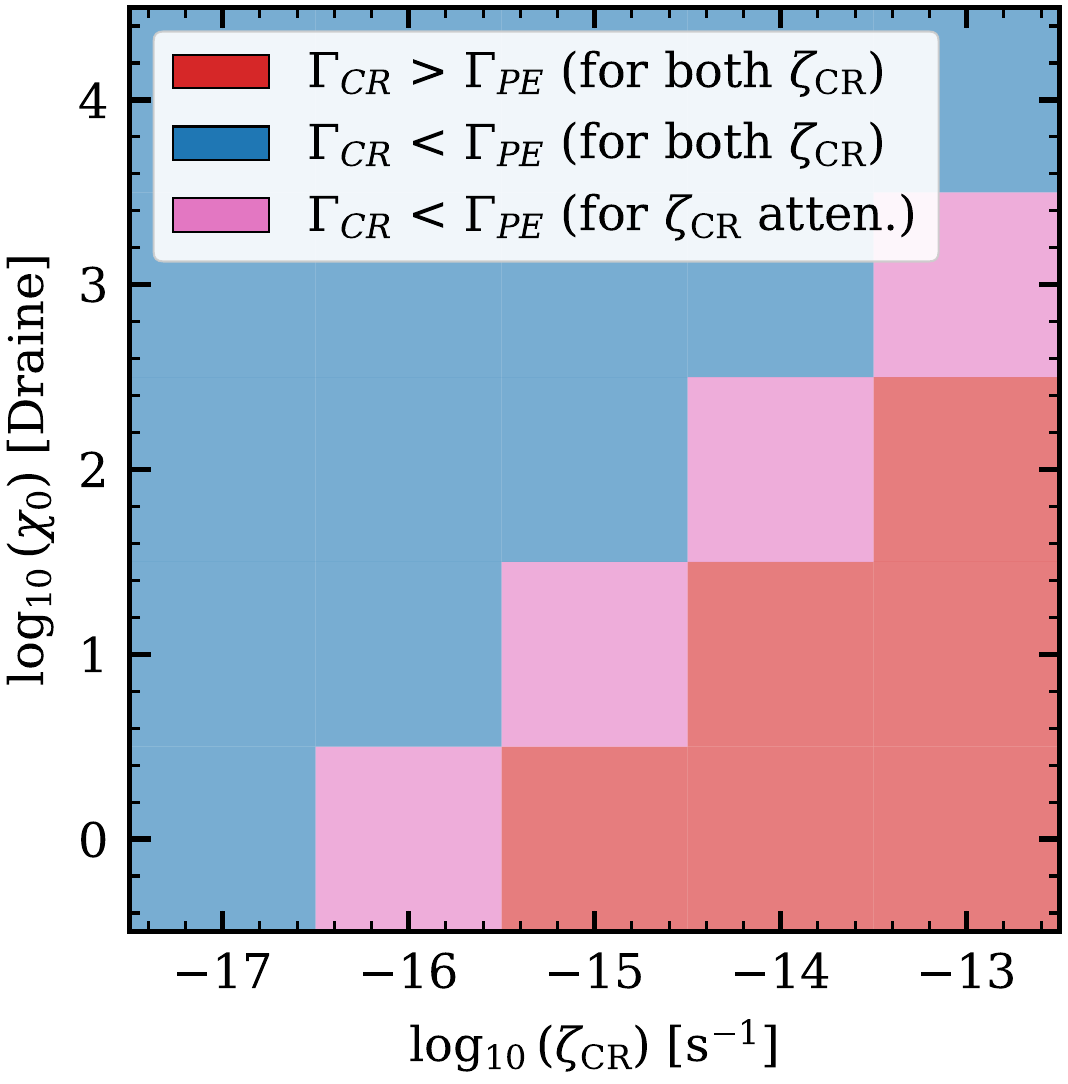}
    \caption{M = 100 M$_{\odot}$, $n_\tot$ = 10$^3$ cm$^{-3}$ (left) and $n_\tot$ = 10$^5$ cm$^{-3}$ (right)}
    \label{fig:heating_processes_a}
\end{subfigure}
\hfill
\begin{subfigure}[t]{0.48\textwidth}
    \centering
    \includegraphics[width=0.48\linewidth]{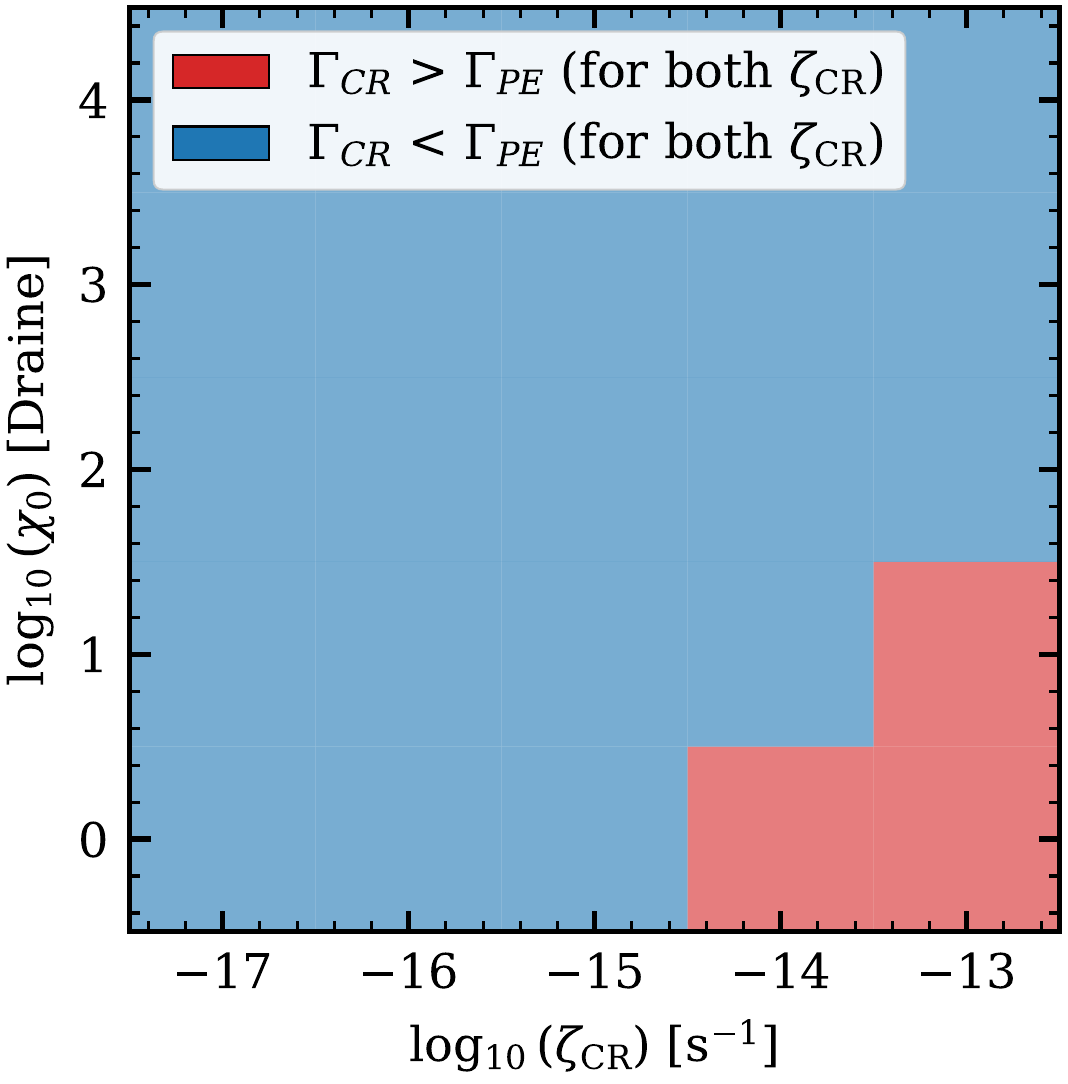}
    \hfill
    \includegraphics[width=0.48\linewidth]{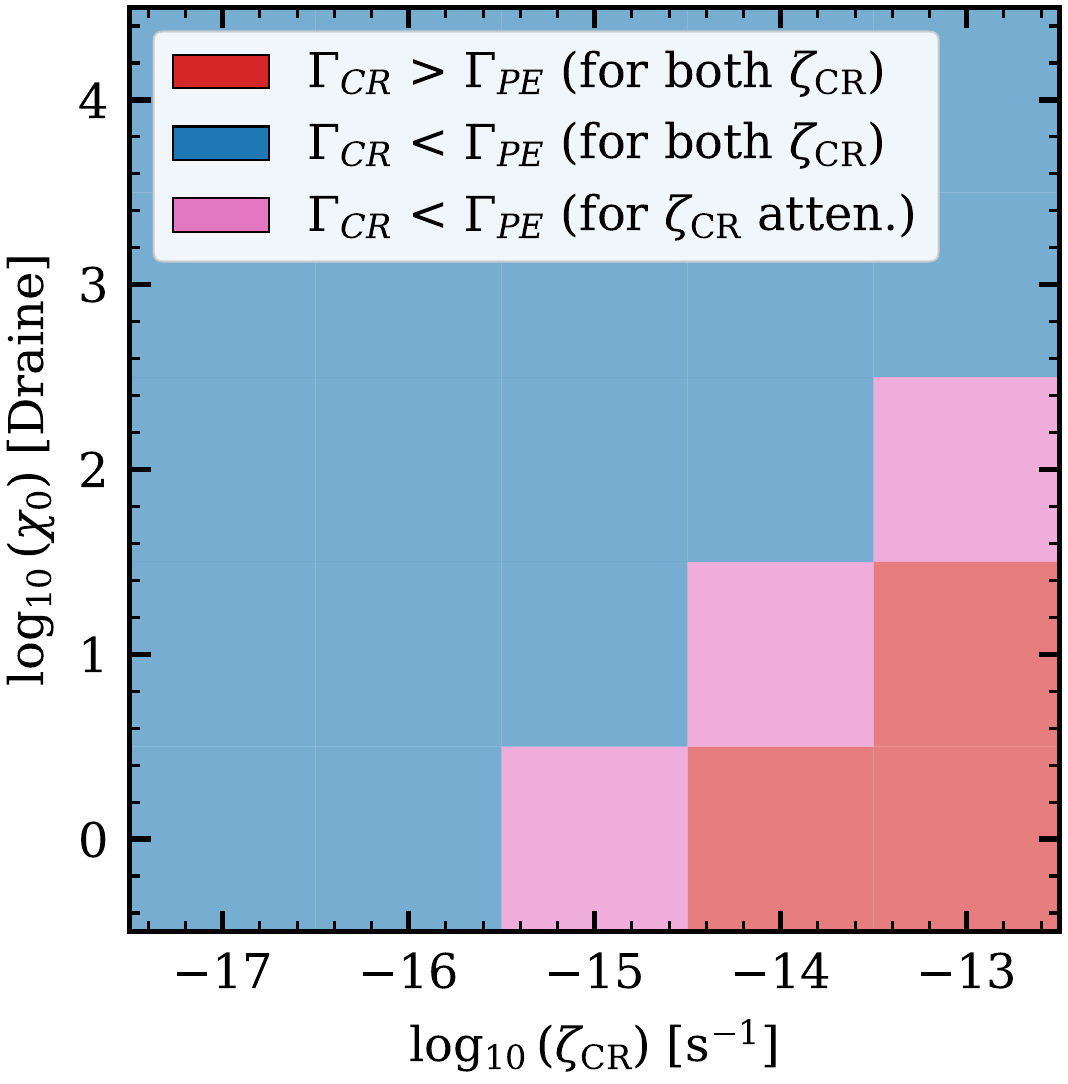}
    \caption{M = 10$^{-2}$ M$_{\odot}$, $n_\tot$ = 10$^3$ cm$^{-3}$ (left) and $n_\tot$ = 10$^5$ cm$^{-3}$ (right)}
    \label{fig:heating_processes_b}
\end{subfigure}
\caption{$\Gamma_\mathrm{CR}$ versus $\Gamma_\mathrm{PE}$ dominance in the $\chi_0$ vs \zetacr\ space. We omit the column for \zetacr\, = 10$^{-18}$ s$^{-1}$ and row of $\chi_0$ = 10$^5$, as $\Gamma_\mathrm{PE}$ dominates and CRs have no effect. The red regions indicate parameter space where $\Gamma_\mathrm{CR}$ dominates $\Gamma_\mathrm{PE}$, while blue shows the opposite. Pink boxes denote the regions for which there was a transition from $\Gamma_\mathrm{CR}$ to $\Gamma_\mathrm{PE}$ due to CR attenuation.}
\label{fig:heating_processes}
\end{figure*}

As shown in Fig. \ref{fig:heating_cooling_mech}, $\Gamma_\mathrm{PE}$ and $\Gamma_\mathrm{CR}$ are the two most important heating mechanisms. Hence, we make a direct comparison of the two. We also note that gas-grain heating ($\Gamma_{gas-grain}$) plays an important role for low-FUV, high-density, and low-CR environments. Figure \ref{fig:heating_processes} shows the dominance of $\Gamma_\mathrm{CR}$ versus $\Gamma_\mathrm{PE}$ in the parameter space of $\chi_0$ versus \zetacr, for the same density values presented in Fig. \ref{fig:heating_cooling_mech}. The red regime indicates where $\Gamma_\mathrm{CR}$ dominates over $\Gamma_\mathrm{PE}$, while blue the opposite. Pink boxes indicate the transition from $\Gamma_\mathrm{CR}$ to $\Gamma_\mathrm{PE}$ due to CR attenuation. Furthermore, to investigate the impact of clump mass on the model, we also include models with the same initial conditions in (n, $\chi_0$, \zetacr), but with a clump mass of M = 10$^{-2}$ M$_{\odot}$, instead of our original M = 100 M$_{\odot}$. Panels a and b of Fig. \ref{fig:heating_processes} indicate the two different mass regimes (M = 100 M$_{\odot}$ (Panel a) and M = 10$^{-2}$ M$_{\odot}$ (Panel b)).

Figure \ref{fig:heating_processes} shows that the impact of CR attenuation in the low-density regime and M = 100 M$_{\odot}$ only affects the low $\chi_0$ models. For values of $\chi_0$ $\gtrsim$ 10$^{2}$, the FUV radiation is high enough to allow $\Gamma_\mathrm{PE}$ to be the dominant mechanism in all regions of the clump, even for the high \zetacr\, regime. For $\chi_0$ = 10$^{2}$, CRs remain marginally the dominant heating mechanism, only for \zetacr\, = 10$^{-13}$ s$^{-1}$, for both constant-CR and attenuated-CR models, as can also be seen from Fig. \ref{fig:heating_cooling_mech}. Higher densities allow the inner parts of the clump to be dense enough, leading to a narrower FUV-illuminated surface that allows CRs to have an impact even for values of  $\chi_0$ $\gtrsim$ 10$^{3}$. CR attenuation affects all regimes where CRs dominated the heating, lowering their impact and allowing $\Gamma_\mathrm{PE}$ to dominate even at high values of \zetacr. For the lower mass regime, CRs have no impact for $n_\tot$ = 10$^3$ $\mathrm{cm}^{-3}$. For the higher density value of $n_\tot$ = 10$^5$ $\mathrm{cm}^{-3}$, the impact is similar to the M = 100 M$_{\odot}$ and $n_\tot$ = 10$^3$ $\mathrm{cm}^{-3}$ clumps, with the only difference that CRs do not dominate for $\chi_0$ = 10$^{2}$. This stems from the fact that lower-mass clumps are smaller in size, extending the FUV-illuminated surfaces to higher values of $A_\mathrm{v}$.

\subsubsection{Cooling mechanisms}
The lower panels of Fig. \ref{fig:heating_cooling_mech} show the cooling contributions in our models. At $n_\tot$ = 10$^{3}$ cm$^{-3}$ both for $\chi_0$ = 1 and $\chi_0$ = 10$^{2}$ the [\cii] 158 $\mathrm{\mu}$m line ($\Lambda_\mathrm{C^+}$) is the dominant coolant for most \zetacr. This is followed by CO ($\Lambda_\mathrm{CO}$), [\ci] ($\Lambda_\mathrm{C_{370}}$), [\oi] 63 $\mathrm{\mu}$m ($\Lambda_\mathrm{O_{63}}$), [Si$^+$] ($\Lambda_\mathrm{Si^+}$), and gas-grain collisions ($\Lambda_\mathrm{gas-grain}$), not necessarily in this order. The main difference in the dominant coolant mechanisms in this density regime is due to the difference in the gas temperature for a particular \zetacr\, value. Higher values of \zetacr\, heat the gas sufficiently to higher $T_\mathrm{gas}$ values, allowing lines such as [\oi] 63 to contribute to cooling. For \zetacr\, > 10$^{-14}$ s$^{-1}$, $\Lambda_\mathrm{C^+}$ and $\Lambda_\mathrm{O_{63}}$ relative contribution reverses, with $\Lambda_\mathrm{O_{63}}$ becoming the dominant coolant mechanism. This transition reflects the different excitation requirements of these lines. Due to its relatively low excitation energy ($T_\mathrm{ex}$ = 91 K) and critical density ($n_\mathrm{crit}$ $\sim$3$\times$10$^3$ cm$^{-3}$), the [\cii] 158 $\mathrm{\mu}$m line efficiently cools diffuse and moderately warm gas of $\sim$100 K. In contrast, the [\oi] 63 $\mathrm{\mu}$m transition requires substantially higher temperatures ($T_\mathrm{ex}$ = 228 K) and densities ($n_\mathrm{crit}$ $\sim$5$\times$10$^5$ cm$^{-3}$) to become collisionally excited. At $n_\tot$ = 10$^{5}$ cm$^{-3}$, for $\chi_0$ = 1 $\Lambda_\mathrm{CO}$ dominates for \zetacr\, $\lesssim$ 10$^{-16}$ s$^{-1}$. Due to high densities and low radiation fields, most carbon is in CO, thus significantly reducing the $\Lambda_\mathrm{C^+}$ contribution to cooling. At higher values of \zetacr\, (10$^{-16}$ s$^{-1}$ $\lesssim$ \zetacr\, $\lesssim$ 10$^{-15}$ s$^{-1}$), CR-induced dissociation of CO reduces its abundance \citep{Bisbas2017}, reducing its contribution to cooling, but still maintaining it as an important coolant. As the \zetacr\, increases, $\Gamma_\mathrm{CR}$ increases the gas temperature sufficiently for the gas-grain to effectively cool the gas. For $\chi_0$ = 1 and \zetacr\, $\gtrsim$ 10$^{-16}$ s$^{-1}$, $\Lambda_\mathrm{gas-grain}$ dominate the cooling, while for $\chi_0$ = 10$^{2}$ the transitions happens for \zetacr\, $\gtrsim$ 10$^{-15}$ s$^{-1}$.

\subsection{Clump-integrated abundance profiles of carbon ladder species (\cii/C/CO)}

\begin{figure}[ht]
\centering

\includegraphics[width=0.49\columnwidth]{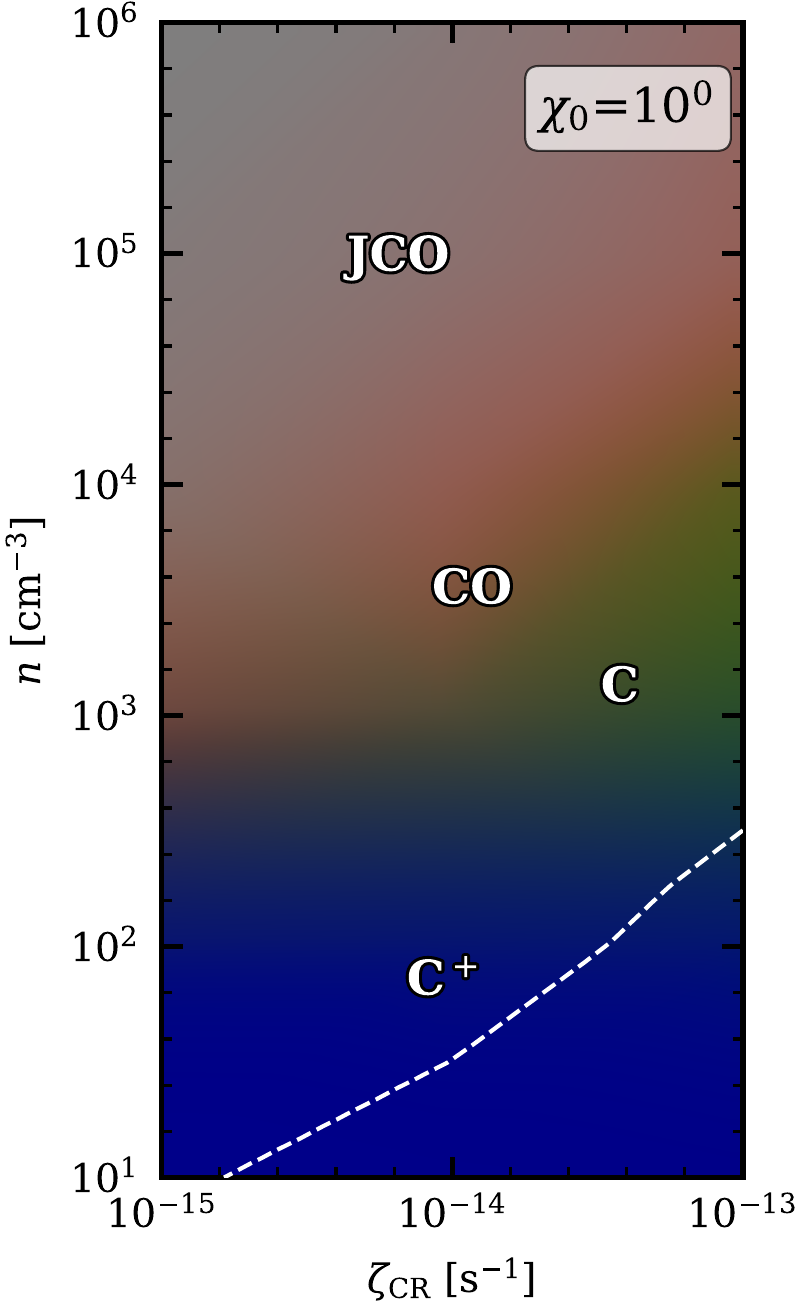}
\includegraphics[width=0.49\columnwidth]{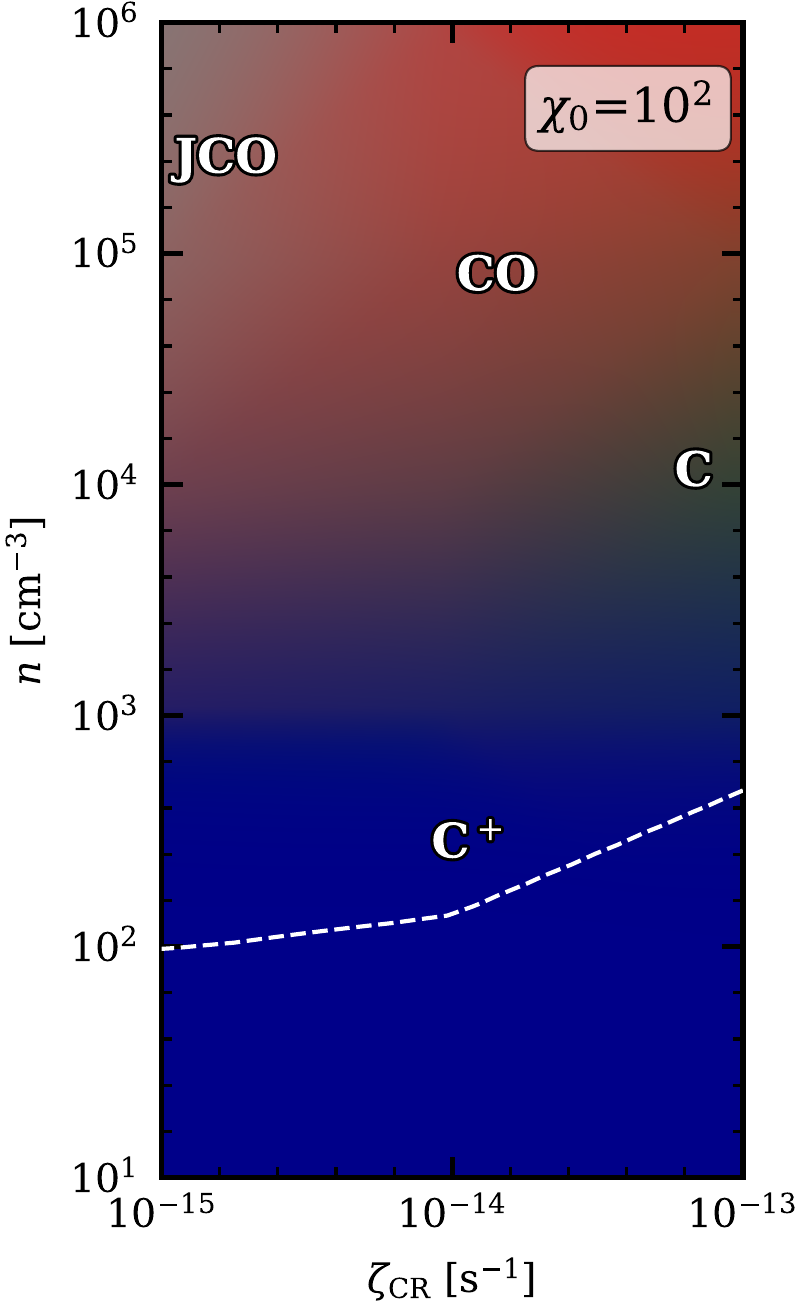}
\caption{Fractional contributions of the carbon ladder species (C$^+$ (blue), C (green), and CO (red)), including JCO (gray), to the total carbon content for two different FUV values ($\chi_0$ = 1, 10$^{2}$, in Draine units) in the constant-CR models ($\alpha$ = 0). The white dashed line represents the \hi-to-\hii\, ($n_\mathrm{H}$ = 2$n_\mathrm{H_2}$) transition.}
\label{fig:sp_phase_diag_main}
\end{figure}

To compare the clump-integrated abundance profiles of selected species in our models, we calculate the spherically integrated abundance of each species as:

\begin{equation}\label{eq:tot_abundance}
    X_\mathrm{sp} = \frac{\int_0^R 4 \pi r^2 n_\mathrm{X} dr}{\int_0^R 4 \pi r^2 n_\mathrm{tot} dr},
\end{equation}
where $n_\mathrm{X}$ represents the density of each species for each radius step. Figure \ref{fig:sp_phase_diag_main} presents the fractional contribution computed as:

\begin{equation}\label{eq:fractional_cont}
    F_\mathrm{sp} = \frac{X_\mathrm{sp}}{X_\mathrm{C^+} + X_\mathrm{C} + X_\mathrm{CO} + X_\mathrm{JCO}},
\end{equation}
of the carbon reservoirs (C$^+$/C/CO), including the ice-CO species JCO,  as a function of $n_\tot$ and \zetacr, for the constant-CR models. Red, green, blue, and gray represent CO, C, C$^+$, and JCO, respectively. The dashed line represents the \hi-to-\hii\, ($n_\mathrm{H}$ = 2$n_\mathrm{H_2}$) transition, for each $\chi_0$ regime. As a result of CO freeze-out inclusion, the gas-phase CO abundance at high densities reflects the balance between gas-phase formation, freeze-out, and desorption processes, rather than saturating at the elemental carbon abundance.

In Fig. \ref{fig:sp_phase_diag_main}, for $\chi_0$ = 1, C$^+$ dominates at low densities, while CO becomes the primary carbon-bearing species for densities $n_\tot$ $\geq$ 10$^{3}$ cm$^{-3}$. Atomic carbon occupies an intermediate layer, which becomes more pronounced at higher ionization rates. As $\chi_0$ increases, the transition boundaries shift, reducing the extent of the C layer, favoring a more extended C$^+$-dominated one. The inclusion of JCO in our chemistry allows the formation of a significant fraction of JCO for densities $\gtrsim$ 2.5 $\times$ 10$^{3}$ cm$^{-3}$ and $\chi_0$ = 1, while for $\chi_0$ = 10$^2$ this transition is limited due to elevated dust temperatures, allowing thermal desorption to bring JCO back to gas-phase CO \citep{rollig2022kosma}. Figure \ref{fig:sp_phase_diag_atten} in Appendix \ref{sec:frac_abund_att} shows the same configuration of parameters, but for the attenuated-CR models.

\subsection{Cosmic-ray versus FUV sensitivity of tracer species}\label{sec:fuv_vs_cr}

\begin{figure*}
\centering
\includegraphics[width = 7.in]{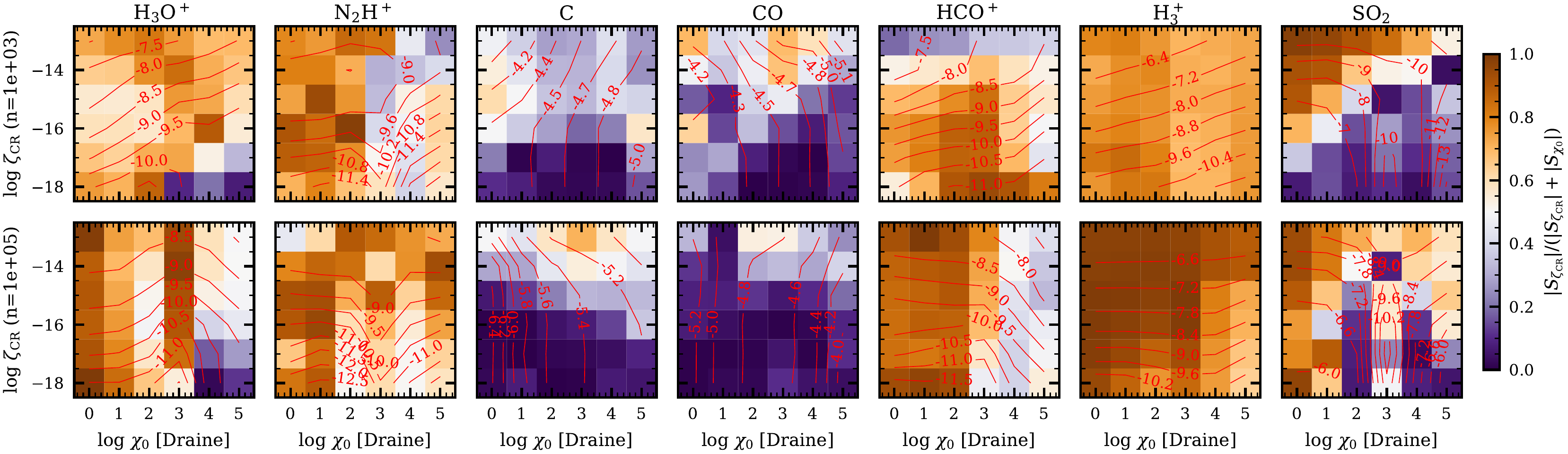}
\caption{Cosmic ray sensitivity fraction, $D$, for H$_3$O$^+$, N$_2$H$^+$, C, CO, HCO$^+$, H$_3^+$, and SO$_2$. Upper panels are for density of $n_\tot$ = 10$^3$\,cm$^{-3}$, while lower panels are for $n_\tot$ = 10$^5$\,cm$^{-3}$. Red contours show the iso-abundance contours marked by the $\log_{10}X_\mathrm{sp}$.}
\label{fig:cr_probe}
\end{figure*}

CRs and FUV photons act through different processes. FUV photons photodissociate and photoionize species in the outer layers of the clump, while CRs ionize H$_2$ throughout its volume and drive the ion--molecule chemistry initiated by H$_3^+$. Their effects on the chemistry nevertheless overlap: both are sources of ions and free electrons, both heat the gas, and both destroy molecules. Consequently, a given $X_\mathrm{sp}$ can result from different combinations of $\chi_0$ and \zetacr. Therefore, an abundance-based (or column density-based) estimate of \zetacr\, is reliable only where the abundance responds more strongly to \zetacr\, than $\chi_0$. To identify these regions, we compare the sensitivity of each species to both parameters first at fixed density.

\begin{figure*}
\centering
\includegraphics[width = 7.in]{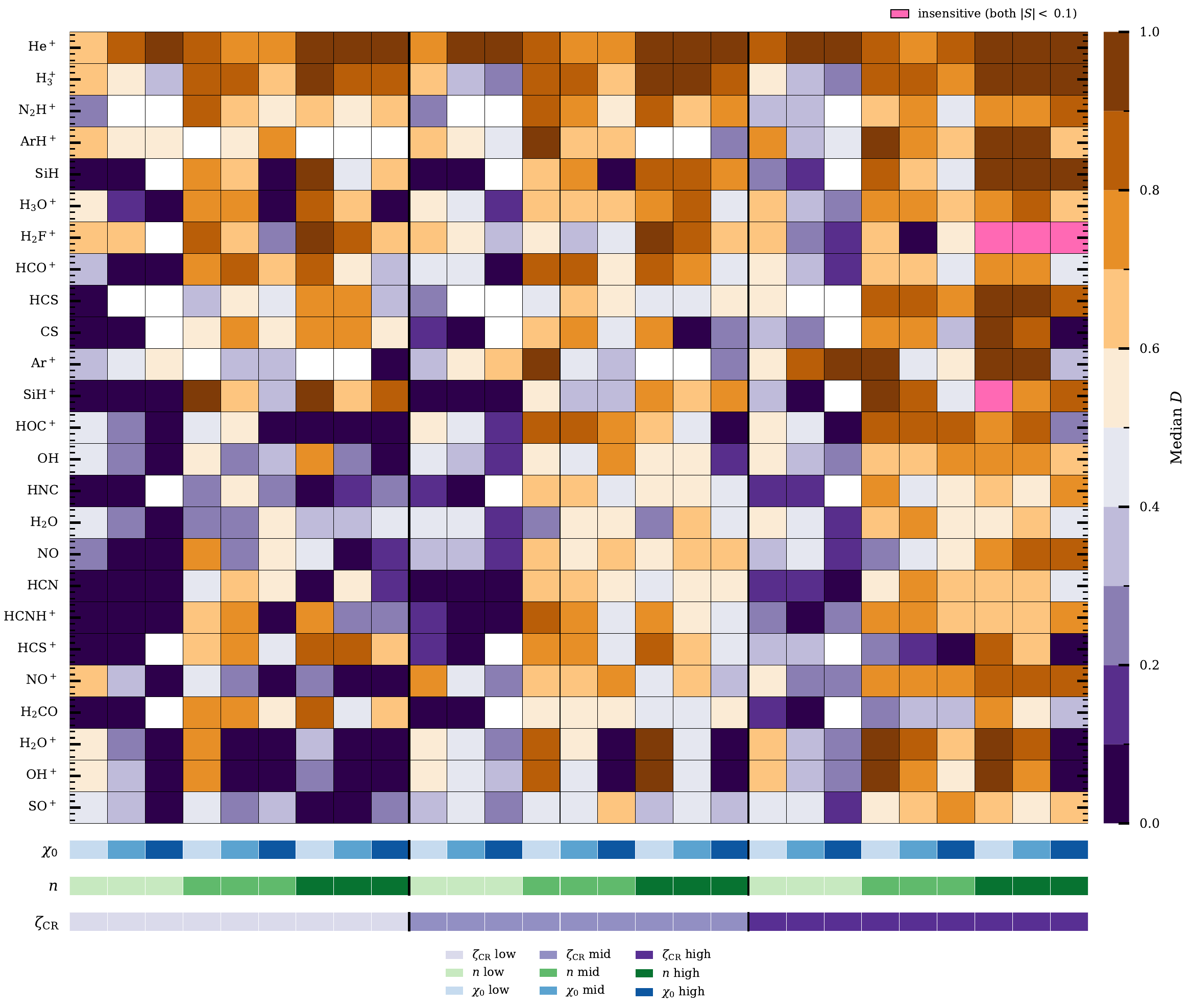}
\caption{Species ranking based on their CR sensitivity fraction $D$. The grid is split into 27 regimes, accounting for all the combinations of low, mid, and high for the three parameters of our grid ($n_\tot$, $\chi_0$, \zetacr).}
\label{fig:cr_dominance_species}
\end{figure*}

To separate the response of each species to CRs from its response to the FUV
field, we compute the local logarithmic sensitivities of $X_\mathrm{sp}$, as calculated using Eq. \ref{eq:tot_abundance}, at fixed densities of $n_\tot$ = 10$^3$ cm$^{-3}$ and $n_\tot$ = 10$^5$ cm$^{-3}$:
\begin{equation} \label{eq:slopes}
  S_{\zeta_\mathrm{CR}} = \frac{\partial \log_{10} X_\mathrm{sp}}{\partial \log_{10} \zeta_\mathrm{CR}}, \qquad
  S_{\chi_0}  = \frac{\partial \log_{10} X_\mathrm{sp}}{\partial \log_{10} \chi_0}.
\end{equation}
The above slopes are evaluated on each density plane of the grid. Between grid nodes, $\log_{10}X_\mathrm{sp}$ is interpolated linearly in the corresponding axes (either $\log_{10} \zeta_\mathrm{CR}$ or $\log_{10} \chi_0$) and the
derivatives are computed as finite differences over a fixed 1 dex window centered on each node. This results in centered differences, $S_\mathrm{i} = (\log X_\mathrm{sp,i+1} - \log X_\mathrm{sp,i-1})/2$, at the interior nodes, and one-sided differences at the grid boundaries. At the boundaries, we truncate the window to the upper-boundary value, so no extrapolation is performed. We define the CR sensitivity fraction as:

\begin{equation}\label{eq:dominance_fraction}
  D = \frac{|S_{\zeta_\mathrm{CR}}|}{|S_{\zeta_\mathrm{CR}}| + |S_{\chi_0}|}
    = \left(1 + \frac{|S_{\chi_0}|}{|S_{\zeta_\mathrm{CR}}|}\right)^{-1},
  \qquad 0 \le D \le 1 .
\end{equation}
Depending on the value of the above equation, a species can respond solely to CRs (i.e., $D$ = 1), solely to FUV changes (i.e., $D$ = 0), or a combination of the two, in between values. We call a region CR-dominated where  $D$ $>$ 0.75 and FUV-dominated where D $<$ 0.25. We note that a threshold of $|S_{\zeta_\mathrm{CR}}|$ $>$ 0.1 is set, to avoid $D$ becoming artificially large if both $|S_{\zeta_\mathrm{CR}}|$ and $|S_{\chi_0}|$ are very small. When this threshold is met, $D$ is set to zero.

Figure \ref{fig:cr_probe} shows the dominance fraction, $D$, as calculated using Eq. \ref{eq:dominance_fraction} for H$_3$O$^+$, N$_2$H$^+$, C, CO, HCO$^+$, H$_3^+$, and SO$_2$. We selected those species as some have been previously used to infer \zetacr\, or have been found sensitive to CR changes \citep{caselli1998ionization, Bisbas2015,Bisbas2017,luo2024new, Indriolo2026}. HCO$^+$ behaves as a CR tracer over a large part of the parameter space. At $n_\tot$ = 10$^5$\,cm$^{-3}$ its abundance is CR-dominated over $\sim$ 60\% of the ($\chi_0$, \zetacr) plane, with nearly horizontal iso-abundance contours for $\chi_0 \leq 10^3$ (red contour lines). At $n_\tot$ = 10$^3$ cm$^{-3}$ the contours tilt (median $D$ $\approx 0.4$), so an uncertainty of 1 dex in $\chi_0$ propagates into $\approx 0.4$ dex in the inferred \zetacr. For \zetacr $\approx 10^{-14}$ s$^{-1}$, the HCO$^+$ abundance has a vertical profile and loses its sensitivity to CRs. Similarly, H$_3^+$ behavior is primarily CR driven. At $n_\mathrm{tot} = 10^3$ cm$^{-3}$ its abundance is CR-driven for 47\% of the ($\chi_0$, \zetacr) plane, but for the remaining 53\% the two effects are intertwined. At $n_\tot$ = 10$^5$\,cm$^{-3}$ its abundance remains CR-driven for 90\% of the ($\chi_0$, \zetacr) plane. N$_2$H$^+$ traces \zetacr\, only at low FUV ($\chi_0 \leq 10^{2}$), at both density regimes. At higher $\chi_0$, the radiation field sets its abundance. The single peak in abundance at $\chi_0$ $\approx$ 10$^3$ is due to CO freeze-out and discussed in Appendix \ref{sec:n2hp_chemistry}. H$_3$O$^+$ and SO$_2$ are only CR-driven for $\sim$ 20\% of the ($\chi_0$, \zetacr) plane, mostly for low $\chi_0$ values. CO and C are essentially insensitive to CRs ($|S_{\zeta_{\mathrm{CR}}}| \lesssim 0.07$ over the grid). Their clump-integrated abundances vary by $\sim$1.1-1.3 dex across the entire ($\chi_0$, \zetacr) space, primarily due to $\chi_0$ variations.

Taking into account the full ($n_\tot$, $\chi_0$, \zetacr) parameter space, we bin our regimes using the method described in Sect. \ref{sec:binning_method}. Figure \ref{fig:cr_dominance_species} shows the first 25 CR-dominated species in those bins. The pink colored boxes denote that $S_{\zeta_\mathrm{CR}}$ $<$ 0.1 in that particular bin for the corresponding species. In those bins , the species is neither CR-dominated nor FUV-dominated. We note here that we selected only species for which spectroscopic data are available in at least one of the two major public molecular spectroscopy databases: the Cologne Database for Molecular Spectroscopy (CDMS) \citep{Muller2001,Muller2005,Endres2016}, and the Jet Propulsion Laboratory (JPL) molecular spectroscopy catalogue \citep{Pickett1998}. Species that do not meet this criterion are excluded from all statistics and tables presented in this work. Each species is ranked according to its $\mathrm{median}(|D|)$ value over all bins. It is evident that even species linked with CRs are still contaminated by FUV radiation in particular environments. An obvious example is H$_3^+$, which in all combinations of low-density bin regimes is severely contaminated by FUV, as was also shown in Fig. \ref{fig:cr_probe}, for density at $n_\tot$ = 10$^3$\,cm$^{-3}$.

\subsection{CR Attenuation effect on the models} \label{sec:attenuation_effect}

\begin{figure*}
\centering
\includegraphics[width = 7.in]{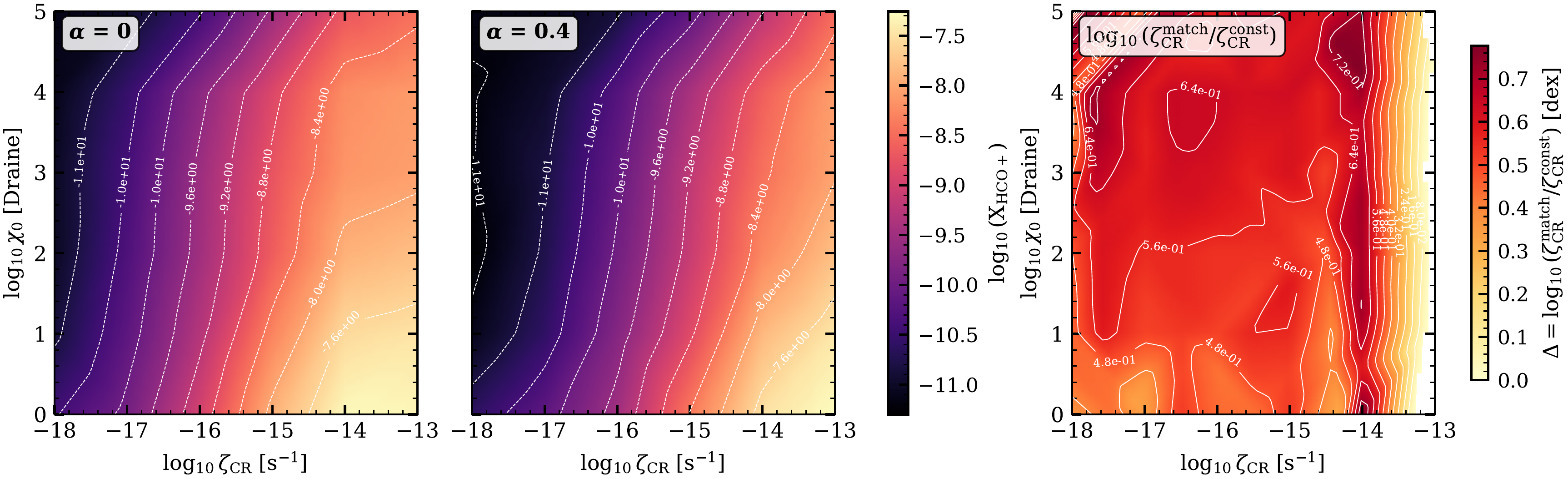}
\caption{Offset diagnostic plots for a density slice of $n_\tot$ = 10$^3$ cm$^{-3}$. \textit{Left}: Total abundance grids for the constant-CR models. \textit{Middle}: Total abundance grids for the attenuated-CR models, adopting an attenuation coefficient of $\alpha$ = 0.4. \textit{Right}: Calculated $\Delta$ parameter based on Eq. \ref{eq:delta_error}, to quantify the offset introduced by CR attenuation.}
\label{fig:delta_triple_plot}
\end{figure*}

\begin{figure*}
\centering
\includegraphics[width = 7.in]{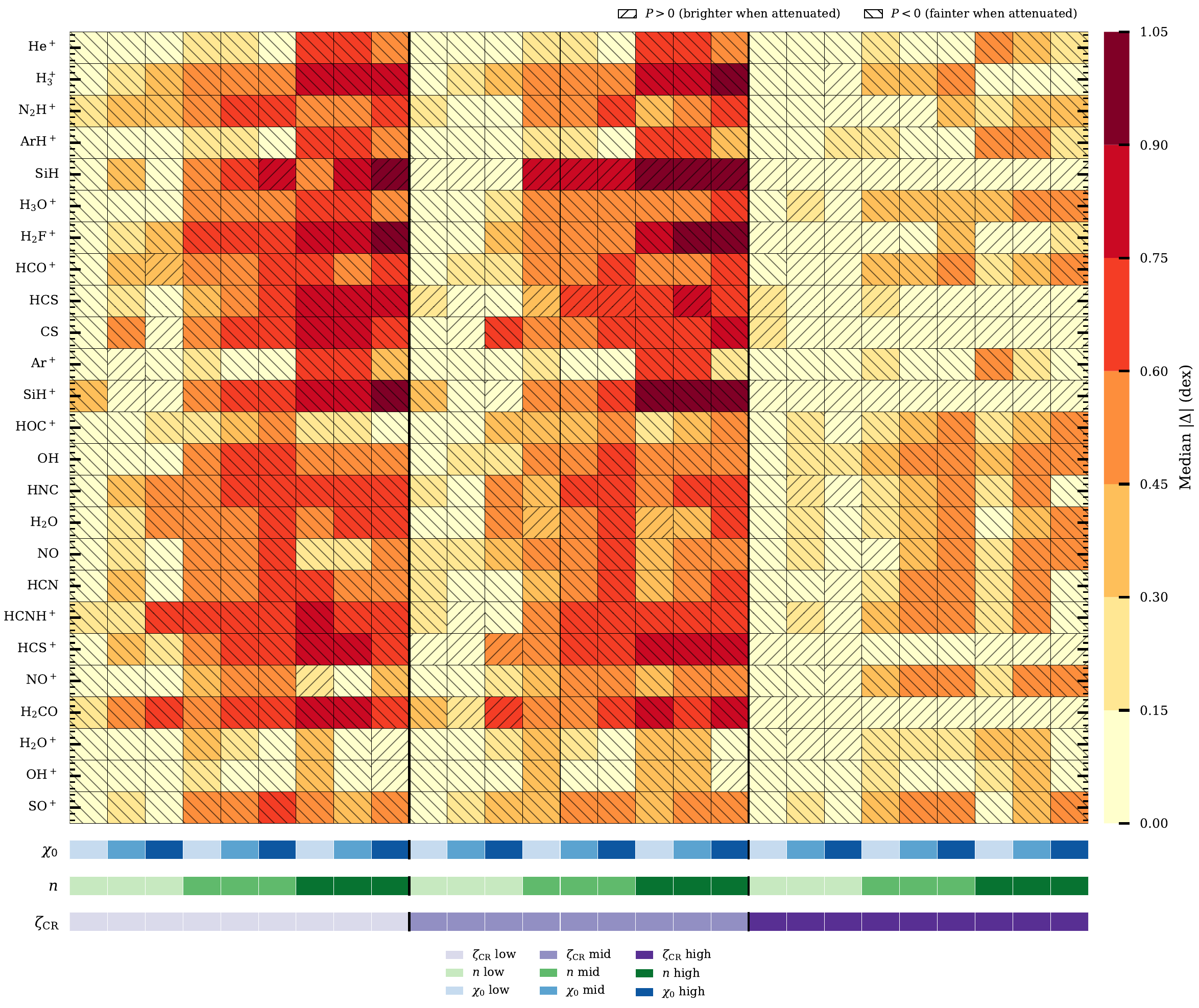}
\caption{Statistical representation of the effect of attenuation for the 25 species presented in Fig. \ref{fig:cr_dominance_species}, across the \kosma\, model grid. The grid is split into 27 regimes, accounting for all the combinations of low, mid, and high for the three parameters of our grid ($n_\tot$, $\chi_0$, \zetacr). Diagonal hatching gives the sign of the median abundance response $P$: rightward hatching (i.e., (\textbackslash\textbackslash): $P$$<$0) denotes a decrease in abundance, while leftward hatching (i.e., (//): $P$$>$0) denotes an increase in abundance.}
\label{fig:species_statistics}
\end{figure*}

Assuming that the effect of CR attenuation is a shift on the \zetacr-axis, we quantify it using the offset: 
\begin{equation}\label{eq:delta_error}
\Delta = \log_{10}\left(\frac{\zeta_\mathrm{CR}^{\,\mathrm{match}}}{\zeta_\mathrm{CR}^{\,\mathrm{const}}}\right),
\end{equation}
where \zetacr$^\mathrm{match}$ is the ionization rate in the attenuated model required to reproduce the clump-integrated abundance predicted by the constant-CR model at \zetacr$^\mathrm{const}$. Assuming that \zetacr$^\mathrm{match}$ $\geq$ \zetacr$^\mathrm{const}$, $\Delta$ indicates that a higher ionization rate is required to compensate for attenuation (see Appendix \ref{sec:shift_method} for more details of the method). Figure \ref{fig:delta_triple_plot} shows a slice of the ($n_\tot$, $\chi_0$, \zetacr) plane at $n_\tot$ = 10$^3$ cm$^{-3}$, for HCO$^+$. The left panel shows the constant-CR models, the middle panel shows the attenuated-CR models, and the right panel shows the resulting $\Delta$. See also Fig. \ref{fig:offset_plot} in Appendix \ref{sec:n2hp_chemistry} for a similar figure for N$_2$H$^+$. Since our native dimensions of the grids are 6$\times$6$\times$6 in the ($n_\tot$, $\chi_0$, \zetacr) space, we interpolate to a 10-times finer grid to allow the change in $\Delta$ to be measurable. Figure \ref{fig:species_statistics} shows the same species as presented in Fig. \ref{fig:cr_dominance_species}, using now the median offset, $\mathrm{median}(|\Delta|)$, to quantify the effect of CR attenuation. In Fig. \ref{fig:species_statistics}, rightwards hatched boxes (i.e., (\textbackslash\textbackslash)) denote regimes for which $P$, as calculated using Eq. \ref{eq:pixel_change}, gave a negative number (i.e., the cell abundance decreased), whereas leftwards hatched boxes (i.e., (//)) denote the opposite (i.e., the cell abundance increased).

All top-ranked species are ones concentrated mostly in the clump center, in regions with elevated column densities. Deeper in the cloud, the main source of C$^+$ and C is through the reaction He$^+$ + CO $\rightarrow$ C$^+$ + O + He \citep{Bisbas2015,Bialy2015}. Lowering the \zetacr\, affects the release of C$^+$ and C in the cloud, which stalls the reaction C$^+$ + H$_2$ $\rightarrow$ CH$_n^+$, which is the main reaction providing the CH$_n^+$ for the formation of all top-ranked affected species \citep[i.e., H$_2$CO, HCNH$^+$, HCN and HNC,][]{Sternberg1995}. Additionally, HCNH$^+$ is formed by reactions of HCN or HNC with H$_3^+$, HCO$^+$, and H$_3$O$^+$, and destroyed by dissociative recombination back to HCN or HNC \citep{Herbst1973,mckee1989,caselli1998ionization,Loison2014}. Both formation and destruction paths are directly linked to \zetacr, one through ion abundances and the latter through electron abundance. In at least 10 environmental bins, the $\mathrm{median}(|\Delta|)$ value of all species is $\gtrsim$ 0.5 dex, leading to a factor of $\sim$ 3 difference in \zetacr\, due to attenuation. HCO$^+$ and N$_2$H$^+$ are also significantly affected by attenuation, with the latter to a lesser degree. They have a $\mathrm{median}(|\Delta|)$ estimate of $\sim$ 0.4 dex (a factor of $\sim$ 2.5) and $\sim$ 0.27 dex (a factor of $\sim$ 1.9), for HCO$^+$ and N$_2$H$^+$, respectively, in the measured \zetacr. CO (not shown in Fig. \ref{fig:species_statistics}) shows negligible change across all regimes ($\mathrm{mean}(|\Delta|)$ $\approx$ 0.09 dex), and it does not rank among the first 25 species. Finally, for all species, the effects of attenuation are more pronounced at lower \zetacr. At higher ionization rates, \zetacr\, remains sufficiently high even after attenuation that the chemistry in high column density regions is less affected.

The effect of attenuation is evident both in the density slices of Fig. \ref{fig:delta_triple_plot} and in the full space of Fig. \ref{fig:species_statistics}, leading to shifts of $\Delta$ $\sim$ 0.1-1 dex. However, this shift is not always uniform, as can be seen in Fig. \ref{fig:delta_triple_plot}, and can only be defined where the abundance responds monotonically and strongly to \zetacr, like in H$_3^+$. For species for which this change is not monotonic, or has a more complicated profile (e.g., N$_2$H$^+$ and SO$_2$), small abundance changes can lead to ambiguous shifts. In these regions, the attenuation effect changes the shape of the abundance response instead of shifting it along the \zetacr\, axis. The regions where this is evident coincide with those where the corresponding species are not CR-dominated.

\section{Discussion}\label{sec:discussion}

\subsection{Impact of FUV radiative transfer and CO freeze-out on $X_\mathrm{sp}$}
Compared to our Fig. \ref{fig:sp_phase_diag_main}, \citet{Gaches2026} find a more extended C layer over a broadly similar region of parameter space. Also, \citet{Roy2026} recently post-processed a simulated dense molecular cloud with \texttt{3D-PDR}, reporting CO abundances approaching the elemental carbon abundance, which they attribute to the absence of CO freeze-out in their chemical network. Despite the large differences between \texttt{3D-PDR} and \kosma, we can discuss two factors that contribute significantly to the observed differences. We also have to consider that \citet{Gaches2026} report the local 1D abundances, while here we have the volumetric average abundance (Eq. \ref{eq:tot_abundance}) of a spherical clump. First, the assumed geometry
(i.e., in our case, 1-dimensional spherical geometry, compared to a 3-dimensional geometry model that is \texttt{3D-PDR}), and secondly, the chemical network. In the \kosma\, setup, the local FUV radiation field is obtained by averaging the attenuation over multiple lines of sight through the cloud, including both absorption and scattering by the adopted grain-size distribution \citep{rollig2022kosma}. In \texttt{3D-PDR}, the local field is instead obtained by summing the attenuated external field along a discrete set of \texttt{HEALPix} rays, without an explicit treatment of dust scattering \citep{Bisbas2012,Zhu2026}. These different treatments of FUV radiative transfer change how deep FUV photons penetrate the cloud, and can lead to variations in the predicted thermal and chemical structure of the PDR, even for the same physical conditions. In addition, our models include CO freezing onto dust grains, whereas the network of \citet{Gaches2026} and \citet{Roy2026} does not, allowing the abundance of CO in the gas-phase to be regulated by the competition between CO formation, freezing, and desorption. Therefore, unlike models without CO freeze-out, our CO abundance does not artificially converge to the elemental carbon abundance at high densities.

\subsection{Implications for cosmic-ray tracers}\label{sec:cr_implication}

Most observational estimates of \zetacr\, derive it using the abundance of two or more species, under an assumed FUV field, and the corresponding relative abundance of the used species with molecular hydrogen. Sect. \ref{sec:fuv_vs_cr} shows that the first assumption is safe only for some species in part of the parameter space, whereas the assumed density also has an impact on the abundance profile.

The simulated three-dimensional parameter space allows us to investigate the effect of attenuation on species used in analytical methods to estimate \zetacr. In particular, \citet{caselli1998ionization} developed a method for estimating \zetacr\, in dense cloud cores using the abundances of HCO$^+$ and CO. They also used a method using DCO$^+$/HCO$^+$, but since \kosma\, does not include deuterated species we do not consider this method in this work. More recently, \citet{luo2024new} introduced an analytical method for estimating \zetacr\, in hot molecular cores based on a combination of HCO$^+$, CO, and N$_2$H$^+$ abundances. Both methods were compared with numerical chemical models. \citet{caselli1998ionization} tested their analytical solution against a pseudo-time-dependent chemical model based on the code originally developed by \citet{Herbst1989}, whereas \citet{luo2024new} used the \texttt{3D-PDR} code \citep{Bisbas2012}.

Figures \ref{fig:cr_probe} and \ref{fig:cr_dominance_species} show that HCO$^+$ and H$_3^{+}$ are among the most robust CR tracers in our grid. HCO$^+$ remains a robust tracer only in FUV-shielded gas ($\chi_0 \lesssim 10^3$), while H$_3^{+}$ behaves as a CR tracer for most of the parameter space. The behavior of HCO$^+$ supports its use in dense cores, where \citet{caselli1998ionization} employed it in their method. At $n_\mathrm{tot} = 10^3$\,cm$^{-3}$ of Fig. \ref{fig:cr_probe}, the iso-abundance contours tilt, so in translucent, high FUV-exposed gas an uncertainty of $\sim$ 1 dex in $\chi_0$ alone propagates into $\approx 0.4$ dex in \zetacr. While H$_3^{+}$ provides a reliable tracer across most of the parameter space, at densities $n_\tot$ $\lesssim$ 10$^3$ cm$^{-3}$ its reliability is limited primarily to the low-FUV regime ($\chi_0 \lesssim 10^2$). Estimates based on N$_2$H$^+$ and H$_2$D$^+$ (i.e., deuterated H$_3^+$) isotopologues have so far targeted cold, dense, well-shielded clumps and cores \citep[e.g.][]{Indriolo2015,Sabatini2020, Sabatini2023, Redaelli2025}, which is the regime our models predict these species to trace \zetacr\, cleanly. Applying the same tracers to clumps exposed to stronger radiation fields, such as those near massive star-forming regions, could lead to FUV and CR effects being indistinguishable, rendering them less reliable.

Figure \ref{fig:cr_probe} and Fig. \ref{fig:extra_slopes} in Appendix \ref{sec:cr_fuv_dominance_3d_extra} show that OH and H$_3$O$^+$ are degenerate between the two parameters over most of the parameter space. On the other hand, OH$^+$ and H$_2$O$^+$ remain robust CR tracers for $n_\tot$ = 10$^5$ cm$^{-3}$ and $\chi_0 \lesssim 10^3$, whereas for $\chi_0$ $\gtrsim$ 10$^3$ their abundance is sensitive primarily to FUV changes. For the low-density regime, OH$^+$ and H$_2$O$^+$ are CR-sensitive only in a small region for $\chi_0$ $\lesssim$ 10$^2$. Both of these tracers were used to trace \zetacr\, in diffuse, low-FUV gas \citep{Neufeld2017}. Using them outside of this regime would lead to ambiguous \zetacr\, measurements. CO and C are insensitive to \zetacr\, in our clump-integrated models.

The new models of \citet{Roy2026} include destruction by atomic O and C in the steady-state balance of \citet{IndrioloMcCall2012}, to test analytical calibrators of \zetacr. They find that the H$_3^+$ calibrator reproduces the numerical abundances with an error below 30\%. In addition, HCO$^+$/CO and C$^+$/CO ratios follow their analytical calibrations in dense gas, as functions of \zetacr/n$_\mathrm{e}$ and \zetacr/$n_\mathrm{H}$, respectively, whereas N$_2$H$^+$/HCO$^+$ shows no clear trend with \zetacr/$n_\mathrm{H}$. Their initial FUV value was deliberately low ($\chi_0$ = 10, in Draine units, but they filtered for points with $\chi_0$ $<$ 0.1, since this is where the analytic chemistry calibrators are defined), to isolate CR-dominated chemistry effects. Our grid quantifies this FUV contribution directly. In the UV-shielded limit our results agree with theirs. H$_3^+$ traces \zetacr\, over most of our parameter space, and HCO$^+$ is CR-dominated for $\chi_0 \lesssim 10^3$. However, at stronger radiation fields, the HCO$^+$ abundance also becomes sensitive to $\chi_0$ (see Sect. \ref{sec:fuv_vs_cr}). Therefore, calibrations validated in shielded gas should not be applied directly to FUV-exposed clumps without accounting for the local radiation field.

\subsubsection{Uncertainty and error propagation}
In the following, we will quantify the uncertainty introduced in the inferred \zetacr\, due to the assumption of an adopted value of $\chi_0^\mathrm{adopt}$. This effect is translated as the FUV-CR degeneracy introduced in Sect. \ref{sec:fuv_vs_cr}. Additionally, we will consider the systematic error introduced by ignoring the CR attenuation.

Based on the analysis presented in Sect. \ref{sec:fuv_vs_cr}, the logarithmic slopes $S_{\chi_0}$ and $S_{\zeta_\mathrm{CR}}$ of the constant-CR models quantify the propagated FUV uncertainty in the inferred \zetacr. Taking into account the true abundance of a species $X_\mathrm{sp}$ produced in ($\chi_0^\mathrm{true}$, \zetacr$^\mathrm{true}$), an observer who adopts
$\chi_0^\mathrm{adopt}$, offset by:
\begin{equation}
  \delta = \log_{10}\frac{\chi_0^\mathrm{adopt}}{\chi_0^\mathrm{true}},
\end{equation}
and fits \zetacr\, on the same constant-CR grid, must recover the observed abundance. The fitted point ($\chi_0^\mathrm{adopt}$, \zetacr$^\mathrm{fit}$) lies on the iso-abundance contour ($\chi_0^\mathrm{true}$, \zetacr$^\mathrm{true}$), and assuming that the error in abundance is zero, the change in abundance should be zero:
\begin{equation}\label{eq:slope_error}
\begin{split}
  \mathrm{d}\log_{10}X_\mathrm{sp} &= S_{\zeta_\mathrm{CR}}\,\mathrm{d}\log_{10}\zeta_\mathrm{CR}
    + S_{\chi_0}\,\mathrm{d}\log_{10}\chi_0 = 0 \;\Rightarrow \\
  \log_{10}\frac{\zeta_\mathrm{CR}^\mathrm{fit}}{\zeta_\mathrm{CR}^\mathrm{true}}
    &\simeq -\frac{S_{\chi_0}}{S_{\zeta_\mathrm{CR}}}\,\delta ,
\end{split}
\end{equation}
to first order in $\delta$, with the slopes evaluated at the true point. In terms of the CR sensitivity fraction $D$, the magnitude of this error is:
\begin{equation}\label{eq:slope_error_D}
  \left|\log_{10}\frac{\zeta_\mathrm{CR}^\mathrm{fit}}{\zeta_\mathrm{CR}^\mathrm{true}}\right|
  \simeq \frac{1-D}{D}\,|\delta|.
\end{equation}
Therefore, the ratio $-\frac{S_{\chi_0}}{S_{\zeta_\mathrm{CR}}}$ is the uncertainty in the inferred $\log_{10}$\zetacr\, per dex of uncertainty in $\chi_0$.

We now consider the systematic error introduced if we ignore CR attenuation. Let us assume again a source with true parameters $(\chi_0^\mathrm{true}$, \zetacr$^\mathrm{true}$) and attenuated CRs, whose abundance $X_\mathrm{sp}$ is fitted with constant-\zetacr\, models at an adopted FUV field $\chi_0^\mathrm{adopt}$ that differs from the true value. If the FUV field was adopted correctly (i.e., $\delta = 0$, see Eq. \ref{eq:slope_error}), the uncertainty due to that factor is negligible. But the systematic error due to ignoring CR attenuation prevails and can be considered as in Eq. \ref{eq:delta_error}.

Equations \ref{eq:delta_error} and \ref{eq:slope_error} define the corresponding systematic error if we ignore attenuation and the uncertainty to the inferred \zetacr\, due to assuming a species as CR-dominated, respectively. For CR-dominated species (i.e., D $>$ 0.75), the FUV-CR degeneracy is small such that  $|S_{\chi_0}/S_{\zeta_\mathrm{CR}}| < 1/3$. For species that are FUV-dominated, or for which the FUV–CR degeneracy is poorly constrained, this term introduces an uncertainty into the final estimate of \zetacr. As a result, an observational estimate of \zetacr\, based on a single species can be considered reliable only when that species is CR-dominated under the density and FUV conditions of the source, and when the attenuation-induced systematic error is sufficiently small. Table \ref{tab:single_species_errors} in Appendix \ref{sec:obs_values_template} provides the values for $\Delta$ and $\frac{S_{\chi_0}}{S_{\zeta_\mathrm{CR}}}$ for selected species, ready to be used for the error calculation. A full table of this analysis will be provided online.

We can directly apply Eqs. \ref{eq:delta_error} and \ref{eq:slope_error} to the analytical methods of \citet{caselli1998ionization} and \citet{luo2024new} introduced above. Since $\Delta$ is logarithmic, $\Delta_{(\mathrm{HCO^+} /\mathrm{CO})}$ = $\Delta_\mathrm{HCO^+}$ - $\Delta_\mathrm{CO}$. CO is almost unaffected by attenuation, meaning  $\Delta_\mathrm{CO}$ $\approx$ 0, and insensitive to \zetacr\, in dense gas. So, the HCO$^+$/CO ratio used by \citet{caselli1998ionization} inherits the response of HCO$^+$. In dense, FUV-shielded cores, where the method is applied, HCO$^+$ is CR-dominated ($D$ $\simeq$ 0.85-0.90, see Table \ref{tab:single_species_errors}), so the FUV uncertainty defined by Eq. \ref{eq:slope_error} is negligible, while the attenuation term is not ($\Delta\simeq$ 0.5-0.6 dex). Therefore, the inferred \zetacr\, reflects the local attenuated CRIR in the HCO$^+$-emitting gas, which is a factor of $\sim$ 3-4 lower than the ambient rate at the clump surface. HCO$^+$ and N$_2$H$^+$ used by \citet{luo2024new}, show comparable offsets ($\Delta\simeq$ 0.4-0.6 dex). But at $\chi_0$ $\gtrsim$ 2$\times$10$^3$, both species have D $<$ 0.75, making both the attenuation systematic error and the FUV-CR degeneracy uncertainty important. In their work, they tested their analytic formula against constant and attenuated models at $\chi_0$ = 1, finding that it recovers the local ionization rate for $N_{\rm H_2}\gtrsim3.4\times10^{21}$ cm$^{-2}$. This is consistent with our results, showing that the measured \zetacr\, is $\Delta\simeq$ 0.4-0.6 dex below the ambient one.

\section{Conclusions}\label{sec:conclusions}

In this study, we computed a wide parameter space grid of ($n_\tot$, $\chi_0$, \zetacr) using \kosma\, \citep{rollig2022kosma} to identify how CR attenuation and different environmental conditions affect the chemistry and the thermal balance of our model clumps, and what those differences imply for commonly used \zetacr\, calibrators. Our main conclusions are summarized below:

\begin{itemize}
    \item Constant-CR models systematically underestimate the ambient cosmic ray ionization rate. In dense, shielded environments, the abundances of commonly used calibrators, such as H$_3^+$, HCO$^+$, and N$_2$H$^+$, trace the local attenuated \zetacr\, rather than the ambient one. Deriving the ambient \zetacr\, using constant-CR models would underestimate its value by $\Delta$ $\approx$ 0.4-0.7 dex, corresponding to a factor of $\sim$ 2.5–5.
    \item Whether a species behaves as a robust CR tracer depends on the environmental conditions. No species in our parameter space behaves as a universal CR calibrator. H$_3^+$ remains a robust CR calibrator for high-density environments, while for densities $n_{\rm tot}$ $\lesssim$ 10$^3$ cm$^{-3}$ it becomes FUV-sensitive. HCO$^+$ remains robust for moderately irradiated gas and higher density environments, and N$_2$H$^+$ traces \zetacr\, only for $\chi_0$ $\lesssim$ 10$^2$. Treating those tracers outside of those regimes can give misleading results for \zetacr, if a correction is not implemented. At $n_{\rm tot}$ = 10$^3$ cm$^{-3}$ an uncertainty in $\chi_0$ of 1 dex can lead to an uncertainty of $\sim$ 0.4 dex in \zetacr, inferred from HCO$^+$. A robust independent observational measurement of environmental conditions is needed before inferring \zetacr.    
    \item Many species do not respond monotonically to \zetacr\, changes  (e.g., SO$_2$ and N$_2$H$^+$ to a lesser degree), so they should be neither used to estimate \zetacr{} nor its attenuation.  For ``well-behaving'' species the analytical expression provided and the tabulated values in Table \ref{tab:single_species_errors} can be used to obtain \zetacr{} and the corresponding shift from attenuation under conditions of uncertain UV fields.
    \item Species formed in regions with elevated hydrogen column densities (e.g., HCN, HNC, HCNH$^+$, SiH, SiH$^+$, H$_2$F$^+$) are the ones that are most affected by CR attenuation. The effect of attenuation on CO and C is only minor, with a mean value across all environments of |$\Delta$| $\approx$ 0.09 dex.
    \item CR attenuation affects the thermal balance in a limited regime. $\Gamma_\mathrm{CR}$ exceeds $\Gamma_\mathrm{PE}$ at low $\chi_0$ and high \zetacr. Attenuation shifts this boundary to higher \zetacr.
\end{itemize}

This analysis shows the importance of environmental dependence when selecting species to measure \zetacr, along with the effect of CR attenuation on them. While species commonly used to infer this value are mostly robust tracers of CRs, they can be misleading if used under environmental conditions where this correlation does not apply. A multi-tracer approach
that combines species with different environmental sensitivities is therefore preferable. In a subsequent work (Topkaras et al. \textit{in prep}), we will use the derived \kosma\, results presented in this work to perform a parameter fit using total integrated intensities. This will both test the robustness of the results we derived here and also use the applied methods to obtain full maps in the ($n_\tot$, $\chi_0$, \zetacr) space.

\begin{acknowledgements}
      Part of this work was supported by the German
      \emph{Deut\-sche For\-schungs\-ge\-mein\-schaft, DFG\/} project
      number Ts~17/2--1. T.T. gratefully acknowledges the Collaborative Research Center 1601 (SFB 1601 subproject A6) funded by the Deutsche Forschungsgemeinschaft (DFG, German Research Foundation) – 500700252.
\end{acknowledgements}

\bibliographystyle{aa}
\bibliography{bibliography}

\begin{appendix}

\section{Surface chemistry species} \label{sec:surface_species}

Table \ref{tab:surf_chemical_species} lists the surface species included in the chemical network used for this work.

\begin{table}[ht]
\caption{Surface chemical species included in our network.}
\begin{tabular}{@{}ccccc@{}}
\toprule
J(H)              & J(H$^{13}$CO)       & J(C$_2$H$_2$)      & J($^{13}$CH$_3$)     & J(H$_2$O)   \\ 
J(H$_2$)          & J($^{13}$CO$_2$)    & J(C$^{13}$CH)      & J($^{13}$CH$_4$)     & J(O$_2$)    \\
J(C)              & J(C$^{18}$O)        & J(C$^{13}$CH$_2$)  & J(C$_2$H)            & J($^{18}$O)   \\
J(CH)             & J($^{13}$C$^{18}$O) & J($^{13}$CH$_3$OH) & J(H$_2$C$^{18}$O)    & J($^{18}$OH) \\
J(CH$_2$)         & J(CO$^{18}$O)       & J(CH$_3$$^{18}$OH) & J($^{13}$CH$_3$O)    & J(H$_2$$^{18}$O) \\
J(CH$_3$)         & J(HC$^{18}$O)       & J(O)               & J(CH$_3$$^{18}$O)    & J(O$^{18}$O)  \\
J(CH$_4$)         & J(H$_2$CO)          & J($^{13}$CO)       & J(NH$_3$)            & J(N)     \\
J($^{13}$C)       & J(CH$_3$O)          & J(OH)              & J(CO$_2$)            & J(N$_2$)    \\
J($^{13}$CH)      & J(CH$_3$OH)         & J(CO)              & J(HCO)               &        \\
J($^{13}$CH$_2$) & J(H$_2$$^{13}$CO)       & J(NH)              & J(NH$_2$)            &        \\ \bottomrule
\end{tabular}
\label{tab:surf_chemical_species}
\end{table}

\section{Elemental abundances}\label{sec:elem_abundance}
Table \ref{tab:elem_abund} shows the elemental abundances of atomic species that the network uses. The elemental abundances of He, C, $^{13}$C, O, N, Si, and S come from \citet{Simon2011}. The fluorine abundance resembles a depleted value, compared to the total proton density, taken from \citet{Neufeld2005,Wakelam2015}. The Ar elemental abundance is the mean value reported by \citet{Wilson1994} for the Orion A molecular cloud. Finally, the $^{18}$O elemental abundance is derived by assuming an isotopic ratio of $^{16}$O/$^{18}$O $\approx500$, with [O]/[H] equal to 0.000447 \citep{Wilson1994,Lodders2009}.

\begin{table}[ht]
\centering
\caption{Initial elemental abundances of species used by \kosma.}
\begin{tabular}{@{}cc@{}}
\toprule
Element  & Abundance (relative to H nuclei)\\ \midrule
Ar       & 3.29 $\times$ $10^{-6}$ \\
C        & 0.000234\\
$^{13}$C & 3.52 $\times$ $10^{-6}$ \\
F        & 6.68 $\times$ $10^{-9}$ \\
He       & 0.0851   \\
N        & 8.32 $\times$ $10^{-5}$ \\
O        & 0.000447 \\
$^{18}$O & 8.93 $\times$ $10^{-7}$ \\
S        & 7.41 $\times$ $10^{-6}$ \\
Si       & 3.17 $\times$ $10^{-6}$ \\ \bottomrule
\end{tabular}%
\label{tab:elem_abund}
\end{table}

\section{PDR Results Interactive Simulation Mapping \& Analysis toolkit (PRISMA)} \label{sec:prisma}
A large fraction of the investigation in this work was done using our PDR Results Interactive Simulation Mapping \& Analysis toolkit (PRISMA). PRISMA is an open-source toolkit with numerous procedures for analyzing PDR data in HDF5 format, with a user-friendly interface (UI). It is independent of the PDR code used. The only requirement is that the code output is in HDF5 format and follows the naming convention established by the ISM database (ISMBD)\footnote{\url{https://ism.obspm.fr}}. A similar tool \footnote{\url{https://github.com/itamos-ism/3D-PDR/tree/main/PDR-studio}} is available in the \texttt{3D-PDR} repository.

\section{Heating and cooling mechanisms for the attenuated model}\label{sec:remaing_heating_cooling}

Besides $\Gamma_\mathrm{CR}$ \citet{Glassgold2012} and $\Gamma_\mathrm{PE}$, \kosma\, includes H$_2$ photodissociation heating ($\Gamma_\mathrm{H_2,ph}$), H$_2$ formation heating ($\Gamma_\mathrm{H_2,grain}$) using a binding energy of 1.49 eV (i.e., one-third of the total H$_2$ binding energy of 4.48 eV), carbon photoionization heating \citep[$\Gamma_\mathrm{C^+}$,][]{Tielens1985}. As mentioned also in the main text, under most conditions, PE-heating is the dominant PDR heating mechanism. Conversely, photoelectrons recombining with dust grains contribute to gas cooling.

The cooling processes include the [{O {\sc i}}] 63 $\mathrm{\mu}$m line ($\Lambda_\mathrm{O_{63}}$), [{O {\sc i}}] 44 $\mathrm{\mu}$m line ($\Lambda_\mathrm{O_{44}}$), [{O {\sc i}}] 146 $\mathrm{\mu}$m line ($\Lambda_\mathrm{O_{146}}$), [\ci] 610 $\mathrm{\mu}$m line ($\Lambda_\mathrm{C_{610}}$), [\ci] 230 $\mathrm{\mu}$m line ($\Lambda_\mathrm{C_{230}}$), [\ci] 370 $\mathrm{\mu}$m line ($\Lambda_\mathrm{C_{370}}$), [\cii] 158 $\mathrm{\mu}$m line ($\Lambda_\mathrm{C^{+}}$), [\siii] 35 $\mathrm{\mu}$m fine-structure transition ( $\Lambda_\mathrm{Si^+}$), the rotational transitions of $^{12}$CO ($\Lambda_\mathrm{CO}$, J=0–49) and $^{13}$CO ($\Lambda_\mathrm{^{13}CO}$, J=0–49), \hi\ Ly$_{\alpha}$ cooling \citep[$\Lambda_\mathrm{Lyman}$,][]{Spitzer1978}, H$_2$O line cooling \citep[$\Lambda_\mathrm{H_2O}$, ][]{Neufeld1987}, OH line cooling ($\Lambda_\mathrm{OH}$), the \oi\ 6300 \AA\ line \citep[$\Lambda_\mathrm{O_{6300}}$, ][]{Bakes1994}, and H$_2$ photodissociation kinetic cooling \citep[$\Lambda_\mathrm{H_2}$, ][]{Lepp1983}. In addition, $\Lambda_\mathrm{H_2, vib}$, and $\Lambda_\mathrm{gras-grain}$ can either heat or cool the gas. Specifically, gas-grain collisions heat the gas when T$_\mathrm{gas}$ < T$_\mathrm{dust}$, while for T$_\mathrm{gas}$ > T$_\mathrm{dust}$ it acts as cooling mechanism.

\begin{figure*}[!]
\centering
\includegraphics[width = 7.3in]{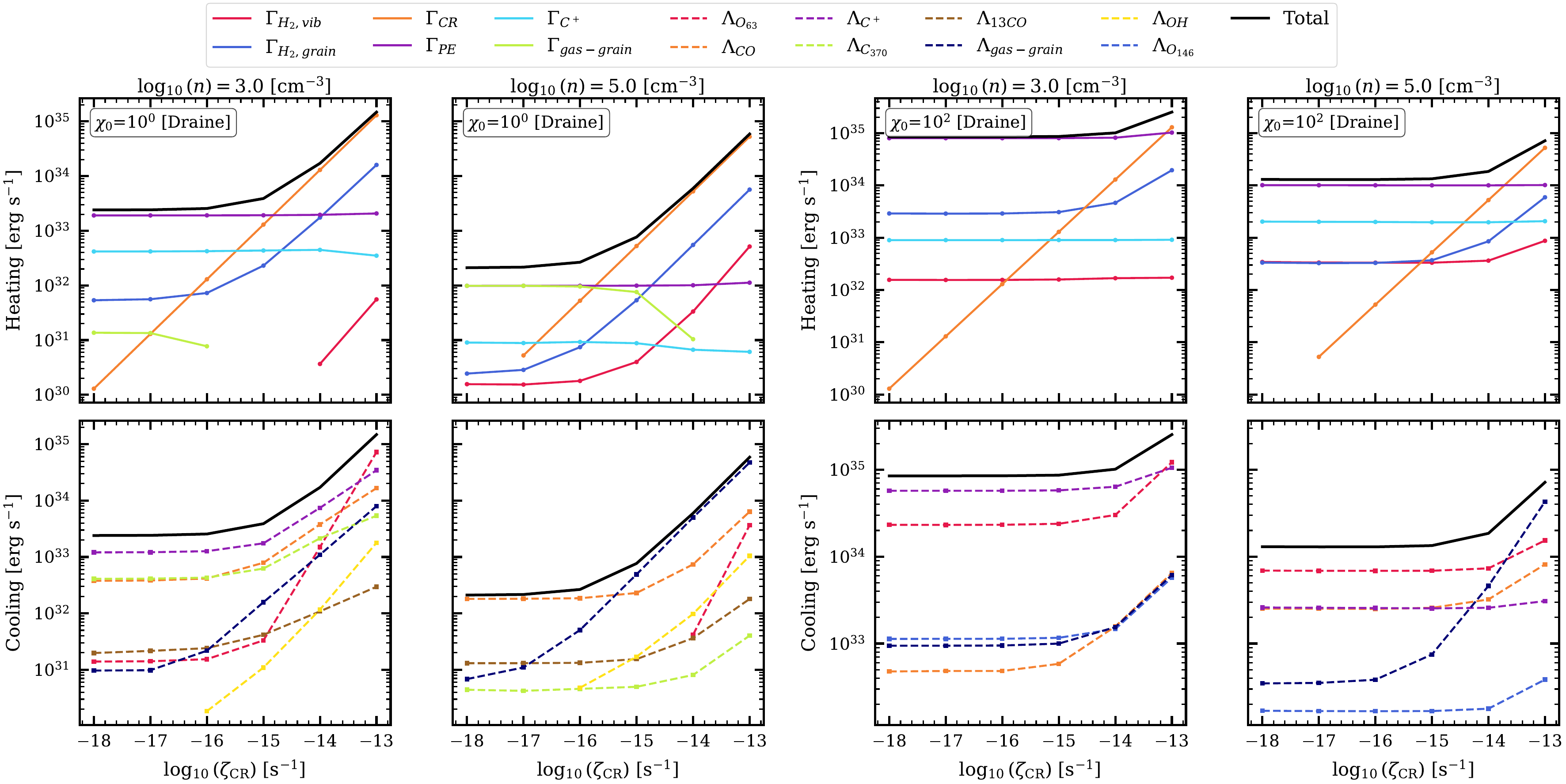}
\caption{Same as Fig \ref{fig:heating_cooling_mech} but for an attenuation coefficient equal to $\alpha$ = 0.4.}
\label{fig:heating_cooling_mech_atten}
\end{figure*}

Similar to Fig. \ref{fig:heating_cooling_mech}, Fig. \ref{fig:heating_cooling_mech_atten} shows the heating and cooling mechanisms, but for an attenuated \zetacr\, using an attenuation coefficient $\alpha$ = 0.4 (see Eq. \ref{eq:pad2009method1}). The attenuation of \zetacr\, affects all heating and cooling mechanisms included in our model. The reduction in $\Gamma_\mathrm{CR}$ is clear across all regimes of n and $\chi_0$, making $\Gamma_\mathrm{PE}$ dominant even at higher \zetacr\, values. $\Gamma_\mathrm{CR}$ remains the dominant heating mechanism for $\chi_0$ = 1, $n_\tot$ = 10$^{3}$ cm$^{-3}$ and \zetacr\, $\geq$ 10$^{-15}$ s$^{-1}$. At $n_\tot$ = 10$^{5}$ cm$^{-3}$ and $\chi_0$ = 1, $\Gamma_\mathrm{CR}$ becomes dominant already at \zetacr\, $\gtrsim$ 10$^{-16}$ s$^{-1}$, as the attenuated FUV radiation field substantially reduces $\Gamma_\mathrm{PE}$ due to dust extinction \citep{Bohlin1978}. In contrast, for the stronger radiation field ($\chi_0$ = 10$^{2}$), $\Gamma_\mathrm{PE}$ remains efficient over a wider range of \zetacr, for both densities, delaying the transition to $\Gamma_\mathrm{CR}$ until the highest values of \zetacr.

The lower panels of Fig. \ref{fig:heating_cooling_mech_atten} show the cooling mechanisms for the attenuated models. At $n_\tot$ = 10$^3$ cm$^{-3}$ and $\chi_0$ = 1, $\Lambda_\mathrm{C^{+}}$ is somewhat enhanced for higher \zetacr\, values. In contrast, $\Lambda_\mathrm{CO}$ remains unaffected, as the total CO abundance does not vary significantly with attenuation. $\Lambda_\mathrm{gas-grain}$ is reduced relative to the constant \zetacr\, models because of the lower gas temperatures. Same $\chi_0$ but for $n_\tot$=10$^5$ cm$^{-3}$ we observe a significant reduction of $\Lambda_\mathrm{gas-grain}$, while $\Lambda_\mathrm{CO}$ is boosted for all \zetacr\, values. This is again attributed to $T_\mathrm{gas}$ < $T_\mathrm{dust}$, which leads the gas-grain mechanism to act as a heating mechanism (mainly for 10$^{-18}$ s$^{-1}$ $\leq$ \zetacr\, $\leq$ 10$^{-15}$ s$^{-1}$) and similar CO abundances for the chosen \zetacr. For $\chi_0$ = 10$^2$, cooling mechanisms are similar compared to Fig. \ref{fig:heating_cooling_mech}, with the only exception that $\Lambda_\mathrm{gas-grain}$ is significantly reduced for \zetacr\, $\geq$ 10$^{-16}$ s$^{-1}$, which leads to a boost of $\Lambda_\mathrm{O_{63}}$, $\Lambda_\mathrm{CO}$, and $\Lambda_\mathrm{C^+}$.

\begin{figure*}[!]
\centering
\includegraphics[width = 7.3in]{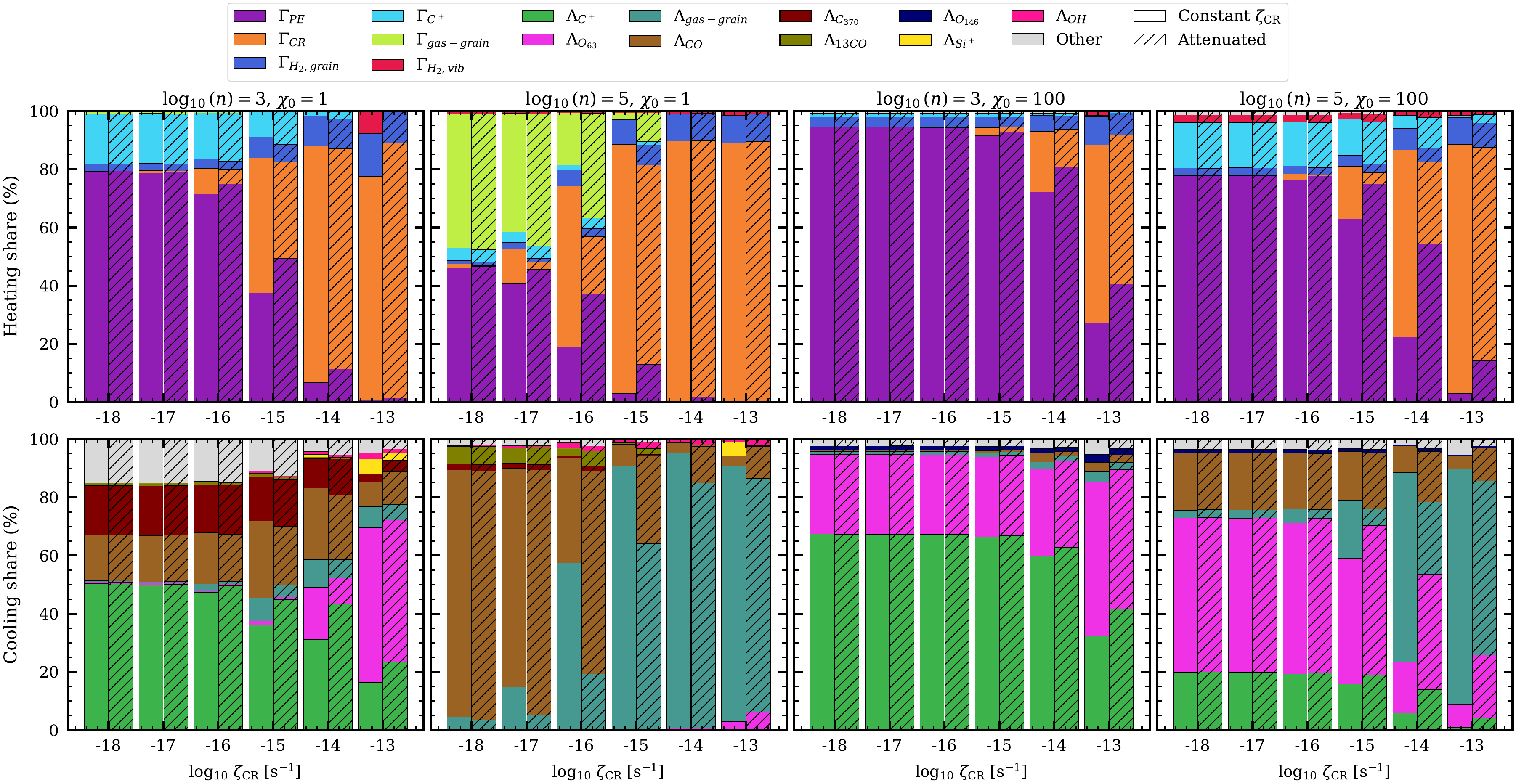}
\caption{Percentage contribution of the dominant heating and cooling mechanisms in the constant- and attenuated-CR models. Percentages are fractions of the clump-integrated total.}
\label{fig:heat_cool_percent}
\end{figure*}

Figure \ref{fig:heat_cool_percent} shows the percentage contributions of the dominant heating and cooling mechanisms included in \kosma, for the constant and attenuated models (hatched bars). The upper panels denote the heating processes and the lower panels denote the cooling. The bars include the percentage contribution to the total for mechanisms with $\gtrsim$10\% dominance. Processes with $\lesssim$10\% dominance (heating: $\Gamma_\mathrm{H_2, ph}$, cooling:  $\Lambda_\mathrm{O_\mathrm{44}}$, $\Lambda_{C_\mathrm{610}}$, $\Lambda_\mathrm{C_{230}}$, $\Lambda_\mathrm{Lyman}$, $\Lambda_\mathrm{H_2O}$, $\Lambda_\mathrm{O_{6300}}$, and $\Lambda_\mathrm{H_2}$) are included in Other (gray).

\section{Attenuated fractional contribution of the carbon ladder species} \label{sec:frac_abund_att}

\begin{figure}[ht]
\centering
\includegraphics[width=0.49\columnwidth]{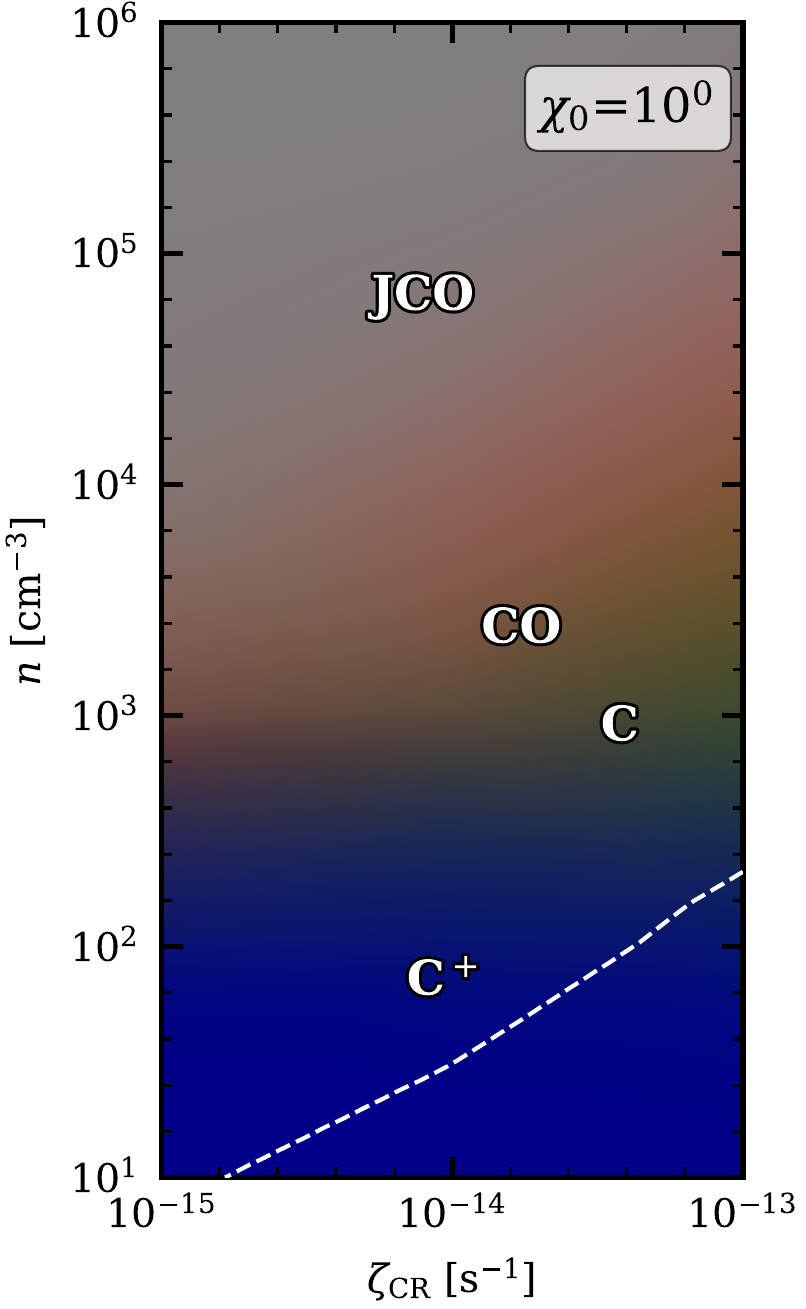}
\includegraphics[width=0.49\columnwidth]{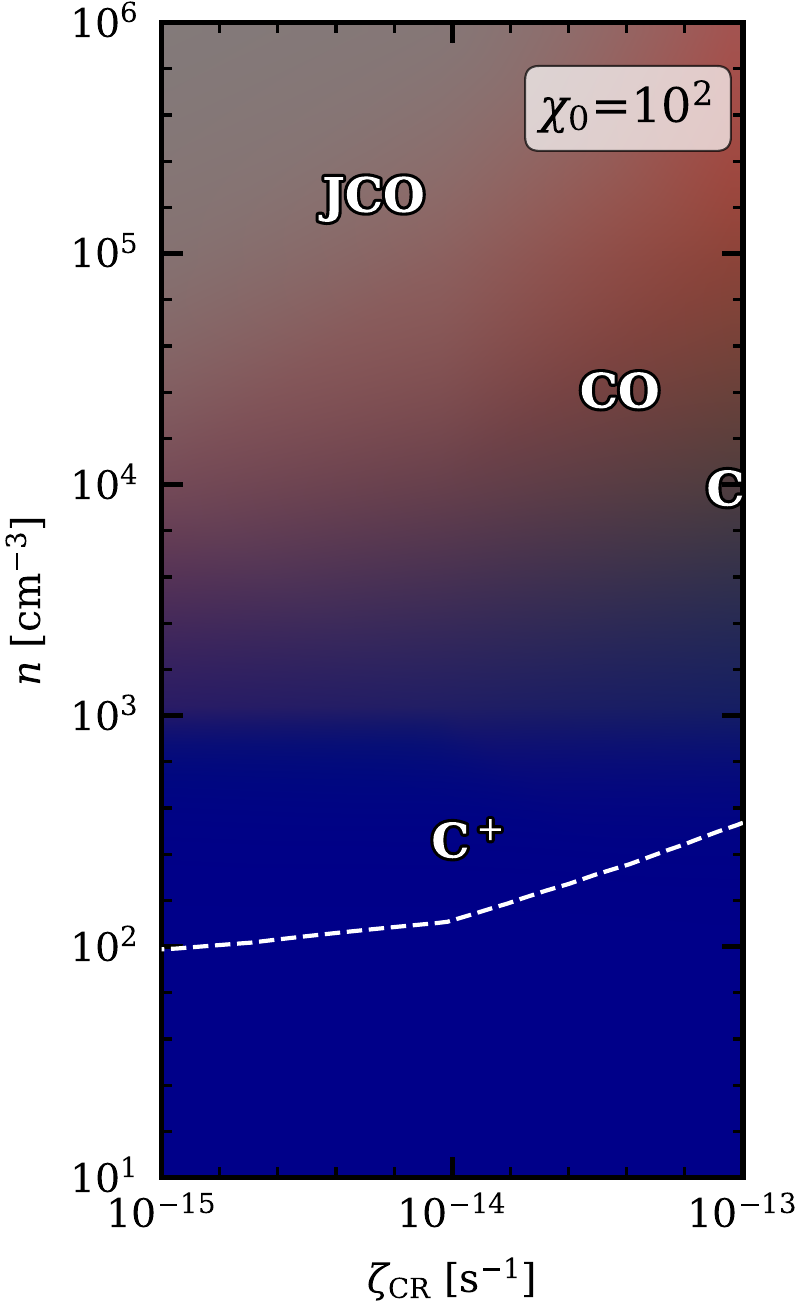}
\caption{Same as Fig. \ref{fig:sp_phase_diag_main}, but for the attenuated models, using $\alpha$ = 0.4.}
\label{fig:sp_phase_diag_atten}
\end{figure}

Figure \ref{fig:sp_phase_diag_atten} shows the fractional contribution of the carbon ladder species (including JCO), as computed using Eq. \ref{eq:fractional_cont}, for the attenuated-CR models. CR attenuation mostly impacts the higher FUV regime where it narrows the C layer, and extends the JCO layer to lower densities. In \kosma, ices desorb thermally, by FUV photodesorption, and by CR-induced desorption through whole-grain heating \citep{Hasegawa1993}. At higher $A_\mathrm{v}$, FUV desorption is suppressed and the dust temperature is low ($T_\mathrm{d}$ $\lesssim$ 10 K) for thermal desorption of CO. This makes CR-induced desorption the dominant desorption channel. As its rate scales linearly with \zetacr, lowering the CR also lowers the CR-induced desorption. Since freeze-out scales with $n_\mathrm{H}$, the ice/gas balance is set by $n_\mathrm{H}$/\zetacr, so a lower \zetacr\, allows CO to freeze out at correspondingly lower densities. Finally, the \hi-to-\hii\, remains unaffected by CR attenuation.

\section{2D grid slices and the effect of CO freeze-out to N$_2$H$^+$} \label{sec:n2hp_chemistry}

\begin{figure*}
\centering
\includegraphics[width = 7.in]{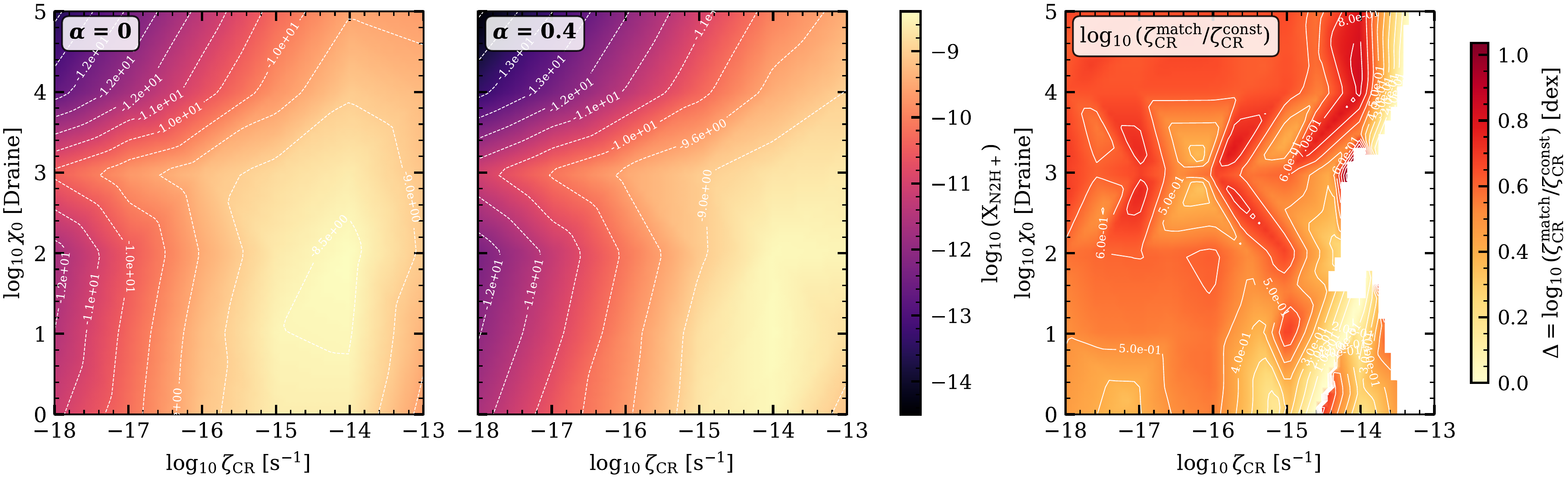}
\caption{Similar to Fig. \ref{fig:delta_triple_plot} but for N$_2$H$^+$.}
\label{fig:offset_plot}
\end{figure*}

Figure \ref{fig:offset_plot} shows a representative 2D slice at $n_\tot$ = 10$^3$ cm$^{-3}$ of $X_\mathrm{N_2H^+}$, for both the constant- and attenuated-CR models (left and middle panels, respectively). Using the method presented in the main text and described in detail in Appendix \ref{sec:shift_method}, we quantify the effect of attenuation on our models.

The inclusion of CO freeze-out in our grids impacts species sensitive to CRs, in particular N$_2$H$^+$. At intermediate FUV of $\chi_0$$\simeq$10$^3$, $X_\mathrm{N_2H^+}$ exhibits a pronounced enhancement (see Fig. \ref{fig:offset_plot}) that cannot be attributed to corresponding increases in the clump-integrated abundances of its main gas-phase formation/destruction partners (N$_2$, H$_3^+$, CO, or electrons). Radial chemical profiles show that this feature arises in a narrow high-$A_\mathrm{v}$ layer at $A_\mathrm{v}$ $\sim$ 5 mag, near the center of the clump, where the dust temperature $T_\mathrm{d}$ $\sim$ 20 K allows for the selective freeze-out of CO onto the grains, while N$_2$ remains mainly in the gas phase. This differential freeze-out is consistent with the higher freeze-out temperature of CO ($T_\mathrm{freeze}$ $\simeq$ 21 K) and N$_2$ ($T_\mathrm{freeze}$$\simeq$ 15 K) \citep{Oberg2005,Bisschop2006}, respectively. With gas-phase CO strongly depleted, the dominant destruction route that is the reaction:
\begin{align}
    \rm CO + N_2H^+ \rightarrow HCO^+ + N_2 \tag{R8} \label{R8},
\end{align}
is suppressed, and N$_2$H$^+$ is instead removed primarily by dissociative recombination with electrons, while formation continues efficiently via:
\begin{align}
    \rm H_3^+ + N_2 \rightarrow N_2H^+ + H_2 \tag{R9} \label{R9}.
\end{align}
At lower $\chi_0$, dust temperatures drop to $T_d$ $\sim$ 15 K, leading both CO and N$_2$ to freeze out. This limits the gas-phase N$_2$ available for N$_2$H$^+$ formation. At higher $\chi_0$$\gtrsim$10$^4$, elevated dust temperatures of $T_\mathrm{d}$ $\sim$ 35 K return CO to the gas phase, restoring efficient proton-transfer destruction of N$_2$H$^+$. The abundance peak at $\chi_0$$\simeq$10$^3$ therefore marks the narrow thermal window of selective CO freeze-out in which N$_2$H$^+$ can accumulate.

\section{Attenuation shift algorithm}\label{sec:shift_method}

CR attenuation changes the clump-integrated abundances predicted by our models. If this change is equivalent to a lower ionization rate, it appears as a shift along the \zetacr\, axis. To quantify it, we compare the constant-CR grid, $X_\mathrm{sp}^\mathrm{const}$, with the attenuated grid, $X_\mathrm{sp}^\mathrm{att}$, where $X_\mathrm{sp}$ is the clump-integrated abundance of species X, for a particular $n_\tot$. At each grid point (\zetacr, $\chi_0$) we define:

\begin{equation}\label{eq:pixel_change}
    P(\zeta_\mathrm{CR}) = \log_{10}\left(\frac{X_\mathrm{sp}^\mathrm{att}(\zeta_\mathrm{CR}, \chi_0)}
    {X_\mathrm{sp}^\mathrm{const}(\zeta_\mathrm{CR}^{\,\mathrm{const}}, \chi_0)}\right).
\end{equation}
Equation \ref{eq:pixel_change} can be used in two ways. (i) Evaluated at the same grid cell, $P(\zeta_{\rm H}^{\,\mathrm{const}})$ quantifies the abundance change due to attenuation (($P$>0): increased abundance, ($P$<0): reduced abundance). (ii) Evaluated along the attenuated grid, $P(\zeta_\mathrm{CR}^{\,\mathrm{att}})$, at fixed $\chi_0$, we find the matched ionization rate, \zetacr$^{\,\mathrm{match}}$, for which $P(\zeta_\mathrm{CR}^{\,\mathrm{match}} = \zeta_\mathrm{CR}^{\,\mathrm{att}}))$ $\simeq$ 0. This value corresponds to the cosmic-ray ionization rate at which the attenuated model reproduces the abundance predicted by the constant-CR model. We then calculate the shift $\Delta$ as:

\begin{equation}\label{eq:offset}
\Delta = \log_{10}\left(\frac{\zeta_\mathrm{CR}^{\,\mathrm{match}}}{\zeta_\mathrm{CR}^{\,\mathrm{const}}}\right),
\end{equation}
where $\Delta>0$ means that a higher ionization rate is needed to recover the reference
abundance after attenuation.

To find \zetacr$^{\,\mathrm{match}}$, we scan the attenuated row starting from \zetacr$^\mathrm{att}$ = \zetacr$^\mathrm{const}$ toward increasing \zetacr. Grid points and the segments between them form one ordered list of candidates, with each point tested before the segment that follows it. Within a segment, the abundance is assumed to vary linearly with $\log_{10}$\zetacr. The first candidate satisfying:

\begin{equation}\label{eq:limit}
\left|X_\mathrm{sp}^\mathrm{att}(\zeta_\mathrm{CR}^\mathrm{match}) - X_\mathrm{sp}^\mathrm{const}(\zeta_\mathrm{CR}^{\,\mathrm{const}})\right|
\leq \delta_{\rm tol}\, X_\mathrm{sp}^\mathrm{const}(\zeta_\mathrm{CR}^{\,\mathrm{const}}),
\end{equation}
is accepted. The tolerance $\delta_{\rm tol}$ = 0.02 only decides whether a grid point is accepted directly. The precision of $\Delta$ is set by the grid spacing ($\sim$0.05 dex in our case, in $\log_{10}$\zetacr). If the row is non-monotonic, this returns the solution nearest \zetacr$^\mathrm{const}$, which it does not need to be the physically relevant one. The profile can depart from the reference value over a wide \zetacr\, interval before crossing it again. Offsets above $\sim$2 dex arise mostly in this regime and should be read as ambiguous matches, not as equivalent ionization-rate shifts. No match exists when (i) the reference abundance lies outside the range spanned by the attenuated row, or (ii) the only solutions lie at \zetacr$^\mathrm{match}$ $<$ \zetacr$^\mathrm{const}$, which we ignore by definition as we assume \zetacr$^\mathrm{match}$ $>$ \zetacr$^\mathrm{const}$. We set $\Delta = 0$ in these cells rather than excluding them, since excluding them would bias the median toward larger offsets.

\section{Complementary slopes of CR-lead species}\label{sec:cr_fuv_dominance_3d_extra}
\begin{figure}[ht]
\centering
\includegraphics[width = 3.4in]{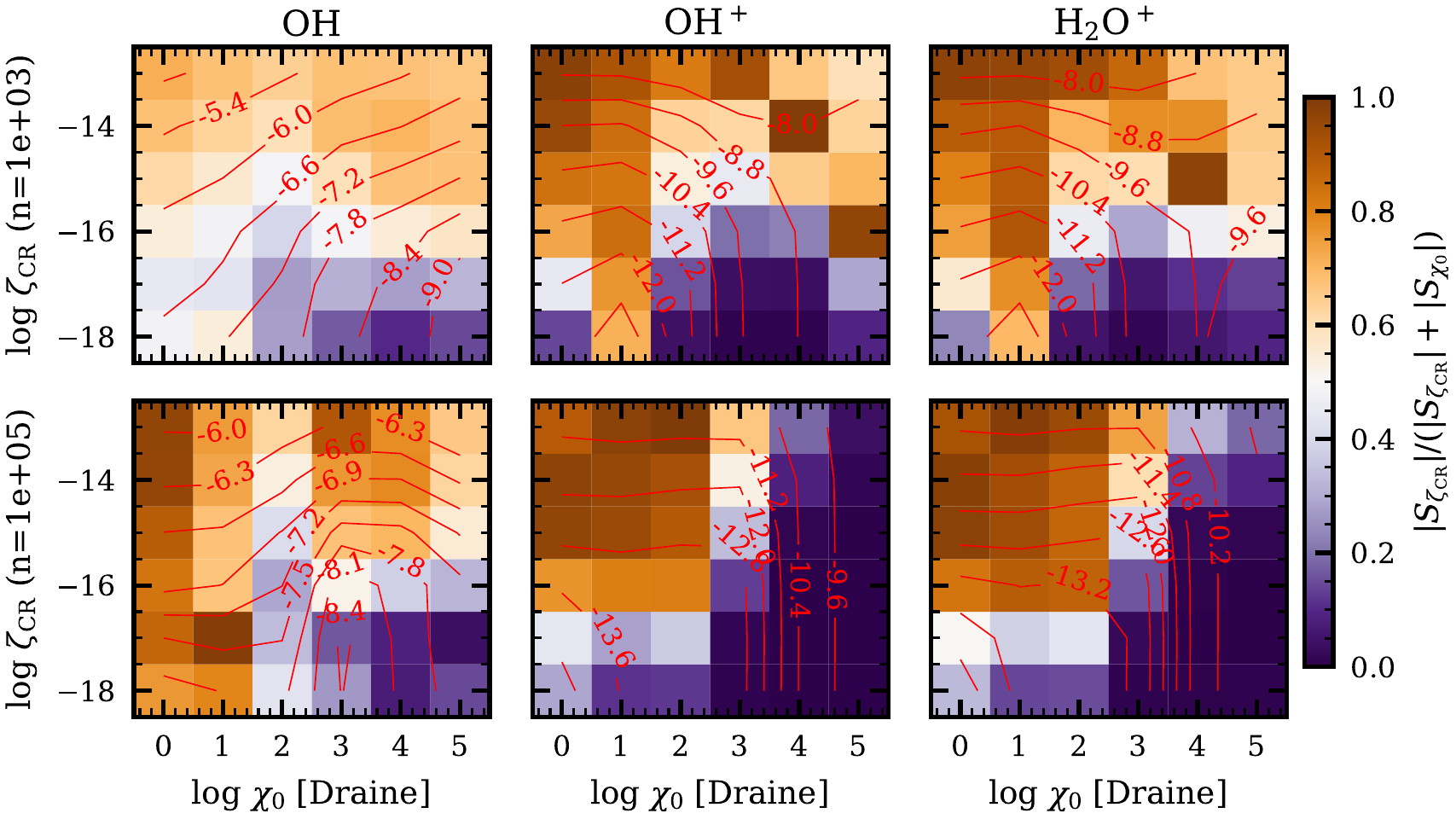}
\caption{Same as Fig. \ref{fig:cr_probe}, but for OH, OH$^+$, and H$_2$O$^+$.}
\label{fig:extra_slopes}
\end{figure}

Figure \ref{fig:extra_slopes} shows the CR dominance fraction, $D$, calculated using Eq. \ref{eq:dominance_fraction}, for OH, OH$^+$, and H$_2$O$^+$. OH$^+$ and H$_2$O$^+$ remain robust CR tracers for $n_\tot$ = 10$^5$ cm$^{-3}$ and $\chi_0 \lesssim 10^3$, whereas for $\chi_0$ $\gtrsim$ 10$^3$ their abundance is sensitive primarily to FUV changes. In the low-density regime, OH$^+$ and H$_2$O$^+$ are CR-sensitive only in a small region for $\chi_0$ $\lesssim$ 10$^2$. OH can only be marginally considered as CR-dominated for high density and $\chi_0$ $\lesssim$ 10$^{2}$. This makes them ambiguous with methods that have used them to trace \zetacr\, in diffuse gas \citep{Neufeld2017}.

\section{Systematic error and uncertainty terms for observations} \label{sec:obs_values_template}

 Table \ref{tab:single_species_errors} shows the median values, over the grid cells of each ($n_\tot$, $\chi_0$, \zetacr) bin, for the attenuation sensitivity $\Delta$, the CR sensitivity fraction $D$, and the inverse relative cosmic ray sensitivity $S_{\chi_0}/S_{\zeta_\mathrm{CR}}$ from our computations for H$_3^+$, HCO$^+$, and N$_2$H$^+$. Both quantities are first evaluated at every cell. The upper and lower offsets are the differences between the 84$^{th}$ and 16$^{th}$ percentiles of that per-cell distribution and its median value. Thus, they describe the spread of the model within the bin. For $\Delta$, cells without a grid match are counted as zero, while for $S_{\chi_0}/S_{\zeta_\mathrm{CR}}$, cells with abundance lower than 10$^{-15}$ cm$^{-3}$ are excluded. We only show the regimes for which the cosmic ray dominance fraction, D, was D $\gtrsim$ 0.75. $f_{\Delta=0}$ denotes the fraction of cells with no match in \zetacr. Regimes with values of $f_{\Delta=0}$ $\gtrsim$ 0.3 should be taken as lower limits on the correction. Finally, the corrected rate is
$\log_{10}$\zetacr$^\mathrm{corr}$ = $\log_{10}$\zetacr$^\mathrm{fit}$ + $\mathrm{median}(\Delta)$,
where the $\mathrm{median}(\Delta)$ of the ($n_\tot$, $\chi_0$, \zetacr) bin is given in Table \ref{tab:single_species_errors}.
The final error in the inferred \zetacr\, and its reported uncertainty are:
\begin{equation}
  \log_{10}\zeta_\mathrm{CR} = \log_{10}\zeta_\mathrm{CR}^\mathrm{fit} + \mathrm{median}(\Delta)
  \pm \sigma_\mathrm{stat}\,(\mathrm{stat})
  \;^{+\sigma_{\mathrm{sys},+}}_{-\sigma_{\mathrm{sys},-}}\,(\mathrm{sys}).
\end{equation}
$\sigma_\mathrm{stat}$ term combines the two independent contributions to the uncertainty, namely 
$\sigma_\mathrm{stat}^2 = \sigma_\mathrm{fit}^2 + (\mathrm{median}(r)^2+\sigma_r^2)\,\sigma_\delta^2$, where $\mathrm{median}(r)=S_{\chi_0}/S_{\zeta_\mathrm{CR}}$, $\sigma_r$ = ($\sigma_{r^+}$ + $\sigma_{r^-}$)/2, and $\sigma_\mathrm{fit}$ is the \zetacr\ measurement error propagated through the fit. $\sigma_\delta$ is the observer's uncertainty on $\chi_0$ in dex. The systematic term, $\sigma_{\mathrm{sys},\pm}$, is the upper and lower offset of
$\Delta$ in the same bin.

\begin{table}[ht]
\centering
\caption{Error and uncertainty terms of Eqs. \ref{eq:delta_error} and \ref{eq:slope_error} for single species estimate of \zetacr\, for H$_3^+$, HCO$^+$ and N$_2$H$^+$.
The columns are the environmental bin ($n_\tot$, $\chi_0$, \zetacr), the attenuation sensitivity $\Delta$, the inverse relative cosmic ray sensitivity $S_{\chi_0}/S_{\zeta_\mathrm{CR}}$, and the CR sensitivity fraction $D$. Regimes ($n_\tot$, $\chi_0$, \zetacr) define the bin introduced in Table \ref{tab:bins}. Here we show only bins for which D $\gtrsim$ 0.75.} 
\label{tab:single_species_errors}
\begin{tabular}{@{}llccc@{}}
\toprule
Species & ($n$, $\chi_0$, $\zeta_\mathrm{CR}$) & $\Delta$ [dex] & $S_{\chi_0}/S_{\zeta_\mathrm{CR}}$ & $D$  \\ \midrule
H3+ & high, high, low & $0.92^{+0.09}_{-0.24}$ & $-0.22^{+0.16}_{-0.26}$ & 0.82  \\
H3+ & high, high, mid & $0.92^{+0.09}_{-0.18}$ & $-0.17^{+0.11}_{-0.19}$ & 0.86  \\
H3+ & high, low, high & $0.28^{+0.58}_{-0.28}$ & $-0.01^{+0.03}_{-0.02}$ & 0.98  \\
H3+ & high, low, low & $0.53^{+0.25}_{-0.17}$ & $0.05^{+0.10}_{-0.06}$ & 0.95  \\
H3+ & high, low, mid & $0.75^{+0.15}_{-0.16}$ & $0.01^{+0.02}_{-0.01}$ & 0.99  \\
H3+ & high, mid, high & $0.31^{+0.61}_{-0.31}$ & $-0.01^{+0.03}_{-0.02}$ & 0.97  \\
H3+ & high, mid, low & $0.64^{+0.25}_{-0.09}$ & $0.16^{+0.11}_{-0.06}$ & 0.86  \\
H3+ & high, mid, mid & $0.79^{+0.16}_{-0.14}$ & $0.02^{+0.00}_{-0.01}$ & 0.98  \\
H3+ & mid, low, high & $0.10^{+0.24}_{-0.10}$ & $-0.20^{+0.14}_{-0.09}$ & 0.84  \\
H3+ & mid, low, low & $0.23^{+0.10}_{-0.05}$ & $-0.11^{+0.16}_{-0.11}$ & 0.88  \\
H3+ & mid, low, mid & $0.28^{+0.12}_{-0.07}$ & $-0.19^{+0.11}_{-0.10}$ & 0.84  \\
H3+ & mid, mid, high & $0.11^{+0.30}_{-0.11}$ & $-0.24^{+0.11}_{-0.15}$ & 0.81  \\
H3+ & mid, mid, low & $0.31^{+0.16}_{-0.11}$ & $-0.16^{+0.27}_{-0.25}$ & 0.81  \\
H3+ & mid, mid, mid & $0.32^{+0.16}_{-0.14}$ & $-0.24^{+0.13}_{-0.14}$ & 0.81  \\
HCO+ & high, low, low & $0.32^{+0.22}_{-0.10}$ & $0.05^{+0.10}_{-0.27}$ & 0.85  \\
HCO+ & high, low, mid & $0.37^{+0.15}_{-0.16}$ & $0.11^{+0.06}_{-0.27}$ & 0.86 \\
HCO+ & high, mid, high & $0.14^{+0.25}_{-0.14}$ & $0.32^{+0.04}_{-0.20}$ & 0.76  \\
HCO+ & mid, low, low & $0.30^{+0.09}_{-0.04}$ & $-0.25^{+0.08}_{-0.16}$ & 0.80  \\
HCO+ & mid, low, mid & $0.27^{+0.02}_{-0.08}$ & $-0.20^{+0.09}_{-0.20}$ & 0.83  \\
HCO+ & mid, mid, low & $0.36^{+0.03}_{-0.03}$ & $-0.08^{+0.56}_{-0.07}$ & 0.87  \\
HCO+ & mid, mid, mid & $0.35^{+0.04}_{-0.04}$ & $-0.11^{+0.39}_{-0.07}$ & 0.86  \\
N2H+ & high, high, high & $0.12^{+0.31}_{-0.12}$ & $0.09^{+0.20}_{-0.09}$ & 0.88  \\
N2H+ & high, low, mid & $0.19^{+0.10}_{-0.07}$ & $0.24^{+0.88}_{-0.18}$ & 0.81  \\
N2H+ & high, mid, high & $0.19^{+0.33}_{-0.19}$ & $-0.09^{+0.26}_{-0.44}$ & 0.77  \\
N2H+ & mid, low, low & $0.24^{+0.02}_{-0.04}$ & $-0.16^{+0.06}_{-0.09}$ & 0.86  \\
N2H+ & mid, low, mid & $0.29^{+0.20}_{-0.05}$ & $0.11^{+0.22}_{-0.21}$ & 0.87  \\ \bottomrule
\end{tabular}%
\end{table}

\section{SO$_2$ and HCN (HNC) chemistry dependence on \zetacr} \label{sec:chemistry_so2}

\begin{figure}[ht]
\centering

\begin{subfigure}{0.48\textwidth}
    \centering
    \includegraphics[width=\linewidth]{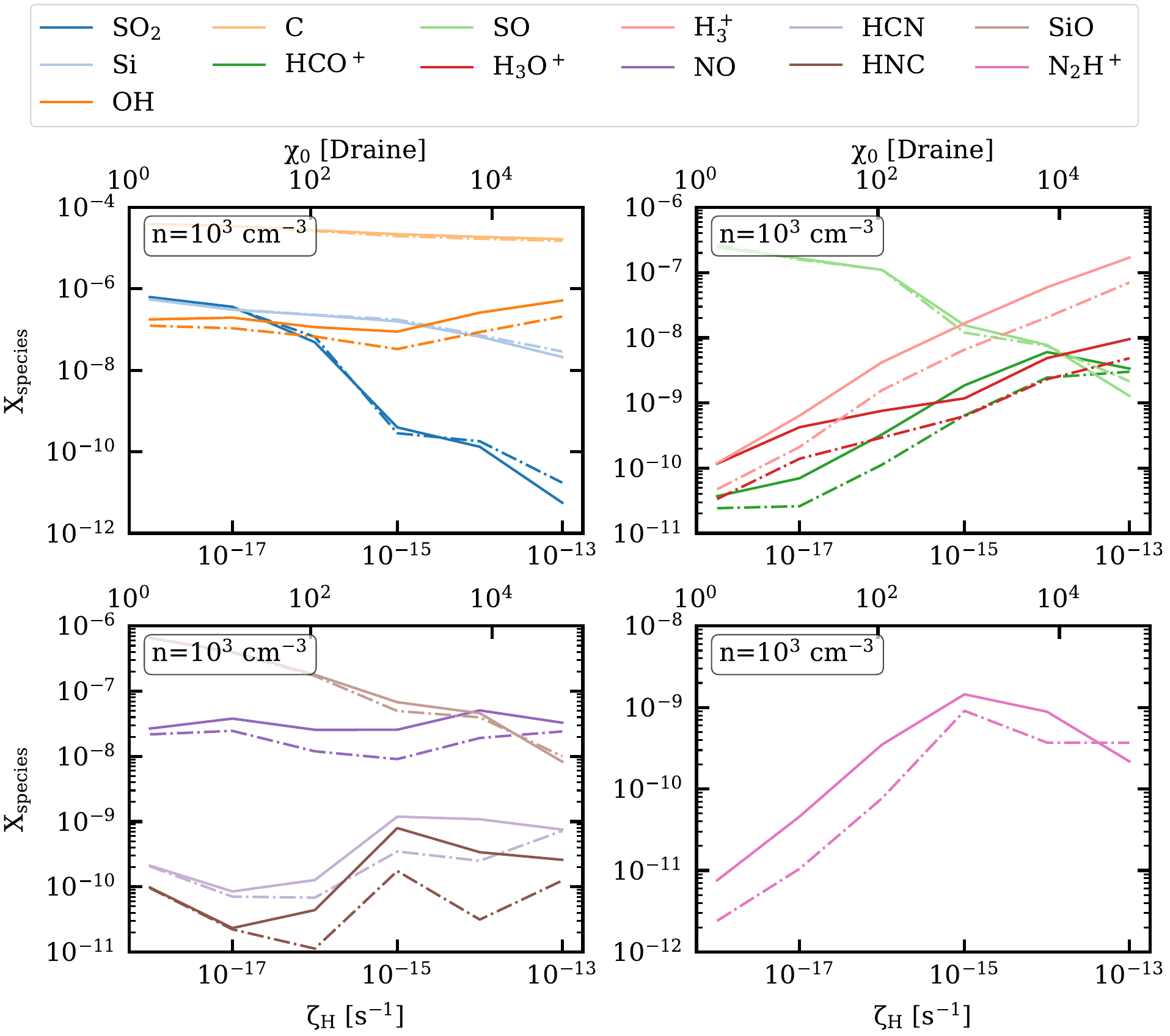}
    \caption{Clump-integrated abundances for $n_\tot$ = 10$^3$ $\mathrm{cm}^{-3}$. }
    \label{fig:n3}
\end{subfigure}
\hfill
\begin{subfigure}{0.48\textwidth}
    \centering
    \includegraphics[width=\linewidth]{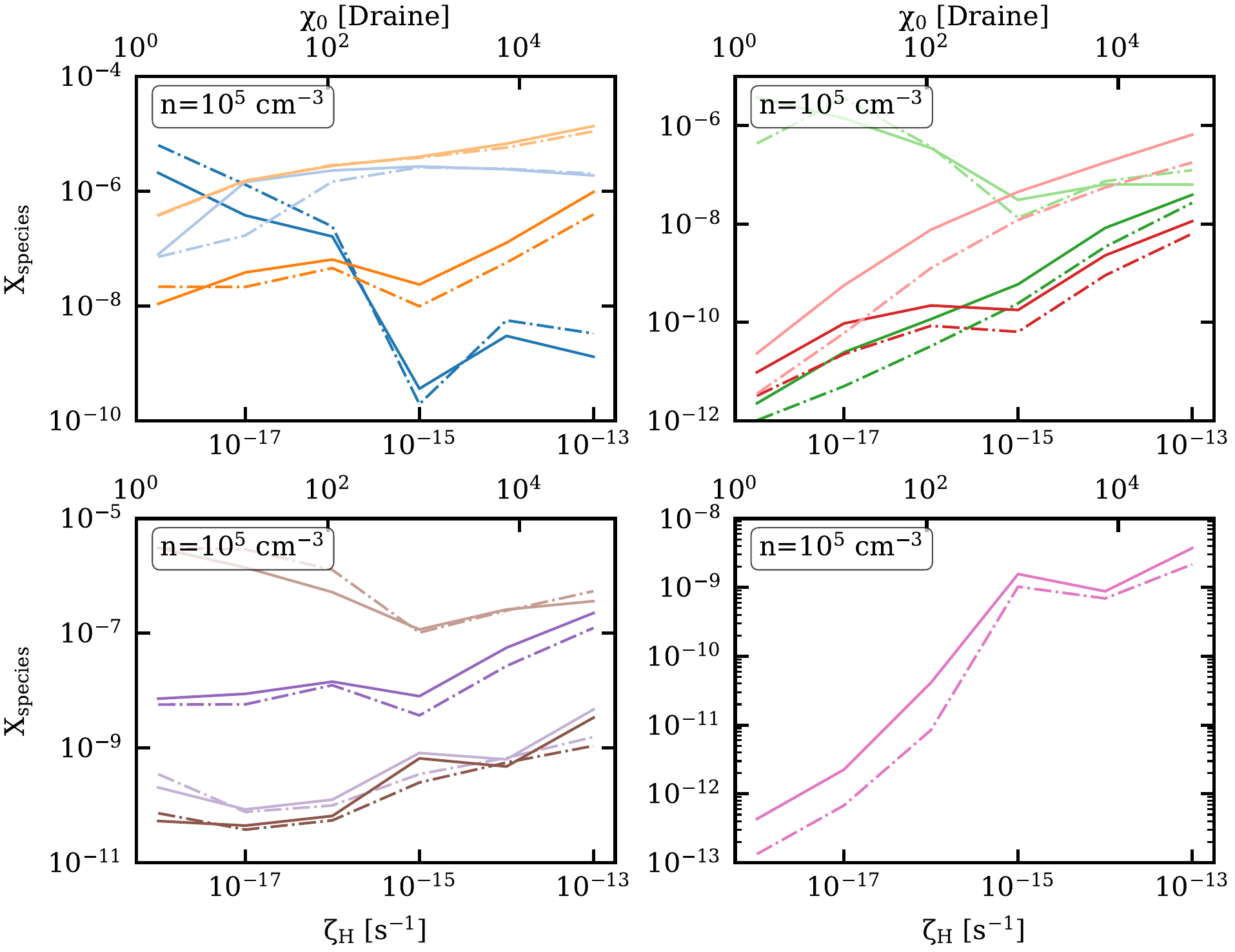}
    \caption{Clump-integrated abundances for $n_\tot$ = 10$^5$ $\mathrm{cm}^{-3}$.}
    \label{fig:n5}
\end{subfigure}
\caption{Clump-integrated abundances of selected species as functions of the FUV radiation field ($\chi_0$ = 1-10$^5$, in Draine units) and the CR ionization rate (\zetacr\, = 10$^{-18}$-10$^{-13}$ s$^{-1}$) for two hydrogen nucleus number densities, $n_\tot$ = 10$^3$ and 10$^5$ $\mathrm{cm}^{-3}$. Solid lines correspond to the constant-CR models, while dash-dotted lines represent the attenuated-CR models.}
\label{fig:sp_tot_abund}
\end{figure}

Figure \ref{fig:sp_tot_abund} shows the $X_\mathrm{sp}$ for selected species that are either directly linked (i.e., produced or destroyed) via CR reactions (e.g., HCO$^+$, N$_2$H$^+$, H$_3$O$^+$, SO) or present a significant difference between the constant and attenuated \zetacr\, models. The upper panel corresponds to $n_\tot$ = 10$^{3}$ cm$^{-3}$, and the lower one to $n_\tot$ = 10$^{5}$ cm$^{-3}$. The solid and dashed-dotted lines represent the constant- and attenuated-CR models, respectively. At density values of 10 cm$^{-3}$ $\leq$ $n_\tot$ $\leq$ 10$^{3}$ cm$^{-3}$, the $X_\mathrm{sp}$ differences between the constant and attenuated CR models are minor. At the density slice of $n_\tot$ = 10$^{3}$ cm$^{-3}$, deviations start to appear between the two models. For the attenuated \zetacr\, models, the $X_\mathrm{sp}$ of almost all selected species decreases relative to the constant \zetacr\, models. We have discussed this effect in Sect. \ref{sec:attenuation_effect} of the main text.

Particular species, such as HCN, HNC, and SO$_2$, are indirectly influenced by CRs, leading to changes in the behavior of $X_\mathrm{sp}$. For HCN and HNC, CRs indirectly influence their chemistry through the abundance of H$_3^+$, which efficiently destroys both species via the ion-neutral reaction:

\begin{align}
    \rm H_3^+ + HCN\,\, (HNC) \rightarrow &\ \rm HCNH^+ + H_2\tag{R2} \label{R2}.
\end{align}
The dominant formation pathway proceeds through the dissociative recombination of HCNH$^+$:
\begin{align}
    \rm HCNH^+ + e^- \rightarrow &\ \rm HCN \,\, (HNC) + H\tag{R3} \label{R3},
\end{align}
whose rate depends on the electron abundance. For both density regimes of Fig. \ref{fig:sp_tot_abund}, the main formation path of electrons for \zetacr\, $\lesssim$ 10$^{-14}$ s$^{-1}$ is primarily through the photoionization of atomic carbon ($\rm C + h\nu \rightarrow  C^+ + e^-$) and for \zetacr\, $\gtrsim$ 10$^{-14}$ s$^{-1}$ through the ionization of molecular hydrogen ($\rm H_2 + p \rightarrow  H_2^+ + e^-$,  where p is a CR proton). Specifically, at $n_\tot$ = 10$^3$ and 10$^5$ $\mathrm{cm}^{-3}$, the $X_\mathrm{H_3^+}$ and $X_\mathrm{C}$ increase by factors of $\sim$50 and $\sim$30, respectively, maintaining chemical steady state requires a corresponding increase in the abundances of HCN and HNC. Consequently, $X_\mathrm{HCN}$ and $X_\mathrm{HNC}$ increase by approximately a factor of $\sim$20.

SO$_2$ is also indirectly influenced by CRs. Its abundance has a turnover at \zetacr\, = 10$^{-15}$ s$^{-1}$ at $n_\tot=10^{5}$ cm$^{-3}$ (lower panels of Fig. \ref{fig:sp_tot_abund}). This behavior is evident in both the constant- and attenuated-CR models. The turnover results from a transition in the dominant destruction pathway, coupled with changes in the OH abundance, which serves as its main formation species. For \zetacr\, = 10$^{-15}$ s$^{-1}$ and $n_\tot$=10$^{5}$ cm$^{-3}$ the main destruction reactions in our chemical network are: 

\begin{align}
    \rm Si + SO_2 \rightarrow &\ \rm SiO + SO\tag{R4} \label{R4}, 
\end{align}
followed by 
\begin{align}
    \rm H_{3}^{+} + SO_2  \rightarrow &\ \rm HSO_{2}^{+} + H_2\tag{R5} \label{R5},
\end{align}
and 
\begin{align}
   \rm C + SO_2 \rightarrow &\ \rm CO + SO\tag{R6} \label{R6}, 
\end{align}
with a contribution to the total rate of 29\%, 24\%, and 22\% for each reaction, respectively. The destruction path changes for \zetacr\, $\geq$ 10$^{-14}$ s$^{-1}$, where the first reaction becomes irrelevant (with a contribution to the total rate of $\approx$ 1\%). This difference comes from the larger relative difference between the total abundance of Si and C, with the latter having a factor of 2 larger abundance for \zetacr\, $\geq$ 10$^{-14}$ s$^{-1}$ compared to \zetacr\, = 10$^{-15}$ s$^{-1}$. For those high \zetacr\, regimes, the CR-induced photo-reaction of SO$_2$ ($\rm SO_2 + h\nu_{CR} \rightarrow  SO + O$) has a contribution of 12\% and 11\% for \zetacr\, = 10$^{-14}$ s$^{-1}$ and \zetacr\, = 10$^{-13}$ s$^{-1}$, respectively.

The main formation path of SO$_2$ for all \zetacr\, is the neutral-neutral reaction with OH:

\begin{align}
    \rm OH + SO  \rightarrow &\ \rm SO_2 + H\tag{R7} \label{R7},
\end{align}
with an almost $\gtrsim$90\% contribution to the total formation rate. For 10$^{-15}$ s$^{-1}$ < \zetacr\, < 10$^{-14}$ s$^{-1}$ and $n_\tot$ = 10$^{5}$ cm$^{-3}$, the $X_\mathrm{OH}$ increases by a factor of $\sim$ 5. Over the same \zetacr\, range, the $X_\mathrm{H_3^+}$ and $X_\mathrm{C}$ increase by factors of $\sim$ 4 and $\sim$ 2, respectively. Although the CR-induced photodissociation reaction discussed above also enhances the destruction of SO$_2$, the increased formation rate dominates, resulting in an overall increase in the SO$_2$ abundance by a factor of $\sim$ 4. Finally, its abundance plateaus and has a small decrease due to the elevated $X_\mathrm{C}$ and the contribution from the CR-induced photodissociation reaction.

\section{Attenuated clump mass fraction and species ranking}
As discussed in Sect. \ref{sec:atten_method}, attenuation effectively begins when $N_\mathrm{eff}$ > $\mathcal{R}$ = 10$^{20}$ cm$^{-2}$. Since \kosma\, uses a density profile to calculate both the clump mass and column densities, some models never reach the attenuation threshold because of their initial conditions. This is the case for models with clump densities of $n_\tot$ $\leq$ 10 cm$^{-3}$. To make statistical predictions about the effect of attenuation on each species, we account for the fraction of the clump mass that is attenuated. We calculate the attenuated clump mass fraction as:

\begin{equation}
    f_\mathrm{att} = \frac{M_\mathrm{att}}{M_\mathrm{clump}} = \frac{\int_0^R 4\pi r^2\, n(r)\,[N_\mathrm{eff}(r) \ge 10^{20}]\,dr}{\int_0^R 4\pi r^2\, n(r)\,dr},
\end{equation}
where $n(r)$ is the density profile as introduced in Eq.\ref{eq:cloud_density}. We find that even for the very low-density models of our grid, 60\% of the regime box presented in Fig. \ref{fig:species_statistics} is attenuated.

\end{appendix}

\end{document}